\documentstyle[psfig,subfigure,lscape]{mn}

\begin{document}
\newcommand{\goodgap}{%
 \hspace{\subfigtopskip}%
 \hspace{\subfigbottomskip}}

\title[The SCUBA Local Universe Galaxy Survey II]{The SCUBA Local
Universe Galaxy Survey II. 450$\mu$m data -- evidence for cold dust in bright IRAS Galaxies.}
\author[L. Dunne, S. A. Eales]
{Loretta Dunne$^{1}$, Stephen A. Eales$^1$\\
$^1$Department of Physics and Astronomy, University of Wales Cardiff,
PO Box 913, Cardiff, CF2 3YB\\ }

\maketitle

\begin{abstract}
This is the second in a series of papers presenting results from the
SCUBA Local Universe Galaxy Survey. In our first paper we provided
850$\mu$m flux densities for 104 galaxies selected from the IRAS
Bright Galaxy Sample and we found that the 60, 100$\mu$m (IRAS) and
850$\mu$m (SCUBA) fluxes could be adequately fitted by emission from
dust at a single temperature. In this paper we present 450$\mu$m data
for the galaxies. With the new data, the spectral energy distributions
of the galaxies can no longer be fitted with an isothermal dust model
-- two temperature components are now required. Using our 450$\mu$m
data and fluxes from the literature, we find that the
450$\mu$m/850$\mu$m flux ratio for the galaxies is remarkably constant
and this holds from objects in which the star formation rate is
similar to our own Galaxy, to ultraluminous infrared galaxies (ULIRGS)
such as Arp 220. The only possible explanation for this is if the dust
emissivity index for all of the galaxies is $\sim 2$ and the cold dust
component has a similar temperature in all galaxies ($T_{c}\sim 20-21$
K). The 60$\mu$m luminosities of the galaxies were found to depend on
both the dust mass and the relative amount of energy in the warm
component, with a tendency for the temperature effects to dominate at
the highest $L_{60}$. The dust masses estimated using the new
temperatures are higher by a factor $\sim 2$ than those determined
previously using a single temperature. This brings the gas-to-dust
ratios of the IRAS galaxies into agreement with those of the Milky Way
and other spiral galaxies which have been intensively studied in the
submm.
\end{abstract}


\section{Introduction}
This paper will present further results from the SCUBA Local Universe
Galaxy Survey (SLUGS), which is the first systematic submillimetre
survey of the local universe. The general aims of the survey are to
provide the statistical measurements (the submillimetre luminosity and
dust-mass functions) necessary for interpreting observations of the
high-redshift universe and to determine the submm properties of dust
in the local universe. The ideal way to carry out the local survey
would be to do a blank field survey of large parts of the sky but this
is currently impractical because the field of view of SCUBA is only
$\sim $2 arcmins. Our approach was to observe galaxies drawn from as
many different complete samples selected in as many different
wavebands as possible, and using accessible volume techniques (Avni \&
Bahcall 1980) to produce unbiased estimators of the luminosity and
dust mass functions. Currently we have completed one sample of 104
objects from the IRAS Bright Galaxy Sample (Soifer et al. 1989)
selected at 60$\mu$m. There is also an an optically selected sample
(86 sources) taken from the CfA redshift survey (Huchra et al. 1983)
for which data analysis is not yet complete (Dunne et al. 2001). This
sample is designed to complement the IRAS sample by probing to lower
submm luminosities, and also to address the possibility that the IRAS
sample is missing a population of galaxies which are dominated by cold
dust (and were therefore undetected by IRAS).

The 850$\mu$m fluxes for the IRAS sample along with first estimates of
the submm luminosity and dust mass functions have been presented in a
previous paper (Dunne et al. 2000, henceforth Paper I). With the data
then available to us (60, 100 and 850$\mu$m fluxes) we were only able
to fit simple isothermal models of the dust spectral energy
distributions (SEDs). While these provided a good empirical
description of that data it was always possible that a more
complicated multi-temperature model would be required in the presence
of more FIR/submm data points. This would have an effect on the
original dust masses we derived under the assumption of a single
temperature, as at 850$\mu$m dust mass is roughly inversely
proportional to the assumed mean dust temperature. The idea of a
multi-temperature model for dust in galaxies is not new -- it was
initially invoked to explain relationships between IRAS observations
and the HI, blue and H$_{\alpha}$ properties of the Milky Way and
other galaxies (Cox, Krugel \& Mezger 1986; de Jong \& Brink 1987;
Lonsdale Persson \& Helou 1987; Boulanger \& Perault 1988;
Rowan-Robinson \& Crawford 1989). In the most common version of these
multi-temperature models, the FIR emission of a galaxy is hypothesised
to originate in (i) a cool `cirrus' component at $T=15-25$ K, arising
from diffuse dust associated with the H{\sc{i}} and heated by the
general interstellar radiation field (ISRF) and (ii) a warm component
at $T>30$ K, from dust near to regions of high-mass star formation and
H{\sc{ii}} regions. The heating source for the warm component is Lyman
alpha photons from young (OB) stars. As the longest passband on the
IRAS satellite was 120$\mu$m and the cold `cirrus' component is
predicted to have a temperature of 15--25 K, only the warm component
would have been detected by IRAS. Using IRAS 60 and 100$\mu$m fluxes
to determine the dust temperature would therefore lead to an
overestimate which, in turn, causes the dust masses to be
underestimated (as at 100$\mu$m $M_{\rm{d}}\propto
T_{\rm{d}}^{-(4+\beta)}$, where $\beta$ is the dust emissivity index
with a value believed to lie between 1 and 2). When the dust masses
estimated using {\em IRAS\/} fluxes are combined with gas mass
measurements, the gas-to-dust ratios are found to be a factor of
$\sim$ 5--10 higher than for our own Galaxy (Devereux \& Young 1990;
Sanders et al. 1991), suggesting that {\em IRAS\/} does not account
for all of the dust mass and strengthening the case for a
two-component model. In order to demonstrate the presence of a cold
component, measurements at wavelengths longer than 100$\mu$m are
required (i.e. where the peak of the cold emission lies). However,
even with a submm measurement at $800-1000\mu$m, the dominance of the
warm component in energy terms ($L\propto T^{4+\beta}$) means that
although most of the mass may reside in cooler dust, the excess
emission produced by it at long wavelengths on the Rayleigh-Jeans tail
is difficult to disentangle from a shallower wavelength dependence of
the dust emissivity (i.e. a lower value of the dust emissivity index
$\beta$) Eales, Wynn-Williams
\& Duncan (1989); Chini et al. (1986).

Due to the increase in spectral coverage brought by instruments such as
COBE, ISO, SCUBA and the IRAM bolometer system, more recent studies
have had access to measurements at multiple submm wavelengths in
addition to IRAS fluxes. These have enabled astronomers to
unambiguously detect cold dust components at 15 $< T_{\rm{d}} <$ 25 K
in several nearby spiral galaxies, including our own (e.g. NGC 891,
NGC 5907, NGC 4656, M51, NGC 3627). Cold dust components have also
been found in more IR luminous/interacting systems such as IC 1623,
NGC 4038/9, NGC 1068, NGC 3079 (Sodroski et al.  1997; Reach et
al. 1995; Alton et al. 1998, 2001; Davies et al. 1999; Frayer et
al. 1999; Papadopoulos \& Seaquist 1999; Dumke et al. 1997; Braine et
al. 1997; Haas et al. 2000; Neininger et al. 1996; Gu\'{e}lin et
al. 1995; Sievers et al. 1994). The 850$\mu$m fluxes we presented in
our first paper did not allow us to investigate a multi-temperature
model because of the uncertainties in the dust emissivity index
$\beta$. We concluded that, in the absence of more submm data points,
a single temperature SED with a $\beta$ of 1.3 provided a good
empirical description of the galaxies in our sample. While we could
not directly test whether there was a colder component present in our
galaxies, we did discuss the implications -- finding that our derived
dust masses would increase by a factor $\sim 2$ and our gas-to-dust
ratios would decrease to a value consistent with those determined for
the Galaxy and other spiral galaxies if there was a colder dust
component present. The extra submm and mm data presented here allow a
better determination of the SEDs of the galaxies in our sample than
was possible in Paper I, and enable us to investigate the presence of
colder dust in these bright IRAS galaxies. We use our improved
knowledge of the dust properties of the galaxies to produce more
reliable dust masses and a revised estimate of the dust mass
function. Finally we compare the derived gas-to-dust ratios of the
galaxies to the values for the Milky Way and other normal spirals.

We will use $H_0=75$ km s$^{-1}$ Mpc$^{-1}$ and $q_0 = 0.5$
throughout.

\section{Observations and Data Reduction}

\subsection{Sample and Observations}
Paper I described our 850$\mu$m observations of a complete sample of
104 objects selected from the IRAS Bright Galaxy Sample (Soifer et
al. 1989). We observed a subset of the BGS with SCUBA on the
JCMT\footnote{The JCMT is operated by the Joint Astronomy Center on
behalf of the UK Particle Physics and Astronomy Research Council, the
Netherlands Organization for scientific Research and the Canadian
National Research Council.}, consisting of all galaxies with
declination from $-10^\circ <\delta< 50^\circ$ and with velocity $>$
1900 km s$^{-1}$, a limit imposed to try to ensure the galaxies fitted
within the SCUBA field of view. The galaxies were observed between
July 1997 and October 1998, using the jiggle-map mode of SCUBA. There
are two arrays on SCUBA, 37 and 91 bolometers for operation at long
(850$\mu$m) and short (450$\mu$m) wavelengths respectively. They
operate simultaneously with a field of view of $\sim$ 2.3 arcmins
(slightly smaller at 450$\mu$m). Typical beam sizes are $\sim 15$
arcsec and 8 arcsec at 850 and 450$\mu$m respectively. Although the
data at 450$\mu$m are taken simultaneously with those at 850$\mu$m,
they are only useful in dry and stable conditions ($\tau_{450} \leq
1.5$) meaning that only a small fraction (17 objects) of our sample
has usable short wavelength data. We supplemented our 450$\mu$m data
for the SLUGS galaxies with two observations from the JCMT archive
which we reduced, and also with submillimetre, millimetre and FIR
fluxes at $\lambda\sim 150 - 1400
\mu$m from the literature.

\subsection{\label{obsS}Data Reduction and Analysis}
The method of data reduction and flux measurement has already been
described in Paper I and also in more detail in Dunne (2000), but
there were some slight differences at 450$\mu$m which will be
presented here. 

The data were reduced using the standard {\sc surf\/}
package (Jenness \& Lightfoot 1998). The atmospheric extinction was
measured using {\em skydips\/} which were performed at frequent
intervals during observing (once every two or three hours). In order
to estimate the zenith sky opacity, $\tau_{\lambda}$, a slab model for
the atmosphere is fitted to the skydip measurements of the sky
brightness temperature. A hot and cold load on the telescope are used
to calibrate the skydip measurements, and the temperatures of these
loads are used in the model fitting. The fits are generally good at
850$\mu$m, but at 450$\mu$m a decent fit can be difficult to produce
with the standard load temperatures. At the suggestion of Wayne
Holland we also performed fits with the following parameters altered:
hot load temperature ($T_{\rm{hot}}$) changed from its default value
to $T_{\rm{amb}}-2$ K, the ambient temperature, $T_{\rm{amb}}$, being
determined from the FITS header of the data; cold load temperature
($T_{\rm{cold}}$) changed to 95 K at 850$\mu$m and 170 K at
450$\mu$m. The combinations of default and new parameters provided
four sets of model fits per skydip.\footnote{Since all this data was
reduced, the JCMT announced that prior to May 2000 the hot and cold
load temperatures used by the skydip software had been inaccurate
(Archibald et al. 2000) and that the standard value of $T_{\rm{hot}}$
had been too high. The recommended changes should not greatly affect
850$\mu$m skydips but can make substantial differences to those at
450$\mu$m depending on the conditions. As the suggested changes are
similar to what we did by varying the hot and cold temperatures, the
$\tau$ derived using the above method are not significantly affected
by this realisation.}  Another estimation of $\tau$ can be made from
the Caltech Submillimetre Observatory (CSO) radiometer measurements at
225 GHz. The CSO radiometer performs a skydip every 10 minutes and
relationships between this so-called `$\tau_{\rm{cso}}$' and the
opacity values at 450/850$\mu$m have been established by the JCMT
staff. There is also a correlation between $\tau_{850}$ and
$\tau_{450}$ which was used to obtain yet another estimate of the
short wavelength opacity, as the 850$\mu$m skydips produced far more
reliable fits than those at 450$\mu$m. All the estimates of
$\tau_{450}$ for each skydip were then averaged (after any obviously bad
fits had been discarded) and that average value used in the extinction
correction.

Fluxes were measured in apertures, chosen using the submillimetre and
optical information as a guide and wherever possible matching that
aperture used to determine the 850$\mu$m flux. The measurement
uncertainties in the fluxes using this method were described in Paper
I, where we found through Monte-Carlo simulations that a
straightforward application of `CCD-type' shot noise greatly
underestimated the true noise in the SCUBA maps. This is because the
SCUBA maps are generally made with 1 arcsec pixels which do not
represent independent regions on the sky. For a CCD the shot noise is
given by $\sigma_{\rm{shot}} = \sigma_{\rm{pix}}\sqrt{N_{\rm{ap}}}$,
where $\sigma_{\rm{pix}}$ is the standard deviation between pixels and
$N_{\rm{ap}}$ is the number of pixels in the object aperture. At
850$\mu$m we found that $\sigma_{\rm{shot}}(850) \approx
8\,\,\sigma_{\rm{pix}}\sqrt{N_{\rm{ap}}}$. We performed the same
Monte-Carlo procedure at 450$\mu$m
and found that now
\[
\sigma_{\rm{shot}}(450) \approx 4.4 \sigma_{\rm{pix}}\sqrt{N_{\rm{ap}}}
\]
The distinction between the shot noise corrections at 450 and
850$\mu$m is due to the differences in resolution at the two
wavelengths; the ratio 4.4/8 is equivalent to the ratio of
their beam sizes (8/15). The factors of 4.4 and 8 can be shown to be
related to the size of the correlated regions within the maps as
$\sqrt{N_{\rm{corr}}}$, ($N_{\rm{corr}}$ being the number of
correlated pixels) so that $N_{\rm{corr}}(850) \approx 64$ and
$N_{\rm{corr}}(450) \approx 19$. These correlated regions are roughly
1/4 of the beam area, the reason for this lying in the linear
weighting routines used by the rebinning task in {\sc
surf}\footnote{For this reason the above relationships for shot noise
($\sigma_{\rm{shot}}$) only apply to maps made in jiggle mode with 1
arcsec pixels and using the linear weighting routines}.

Once the object flux had been measured in the aperture, it was
calibrated in units of Jy using a map of a planet (Uranus or Mars) or
the secondary submm calibrator, CRL 618. The same aperture
was used for the calibrator as for the object (paying attention to the
orientation of the aperture relative to the chop throw, as the beam is
elongated along the direction of the chop). The calibrated flux for
the object is then given by
\begin{equation}
\rm{S_{\rm{obj}}}(\rm{Jy}) = \frac{S_{\rm{obj}}(\rm{V})}{S_{\rm{cal}}(\rm{V})}\times
S_{\rm{cal}}(\rm{Jy})  \label{fluxE}
\end{equation}
where $\rm{S_{cal}(Jy)}$ is the total flux of the calibration object
in Jy which, for planets, was taken from the JCMT {\sc fluxes}
program. It was assumed that CRL 618 had fluxes of $S_{850} = 4.56$ Jy
and $S_{450}=11.2$ Jy (as listed on the JCMT calibration web page).
Note that this is an inherently different method of calibration than
that described in the mapping cookbook (Sandell 1997), which produces
maps in units of Jy beam$^{-1}$ by multiplying by a `gain' or `Flux
Conversion Factor (FCF)'. This quantity (denoted $C_{\lambda}$) is
simply the ratio of the flux of the planet in Jy beam$^{-1}$ divided by
the peak flux of the planet in Volts, as measured on the map.

\subsection{Calibration uncertainty at 450$\mu$m}
Even in the best atmospheric conditions, the zenith sky opacity of the
atmosphere at 450$\mu$m is $\sim 0.6$, and therefore the accuracy of
the final 450$\mu$m fluxes depends critically on the accuracy of the
extinction correction. In this section we describe a detailed
investigation into the accuracy of 450$\mu$m photometry with SCUBA.  
\subsubsection{Extinction Correction}
\begin{figure}
\centerline{\psfig{file=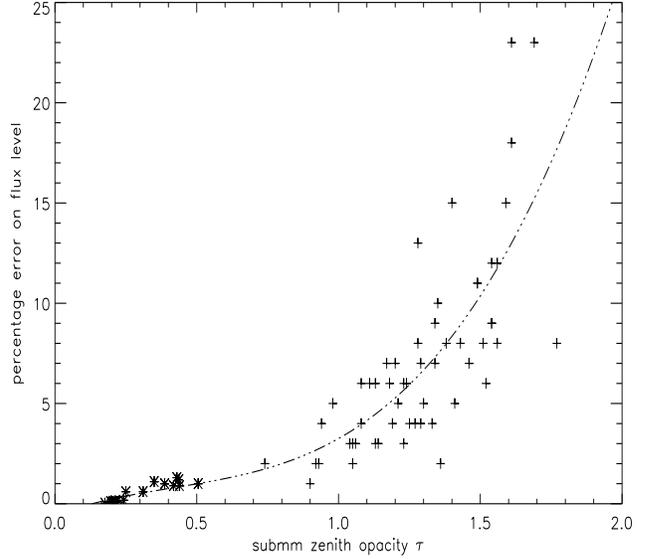,height=8cm,width=9cm}}
\caption[Flux error due to uncertainty in $\tau$]{\label{tauF}
Relative uncertainty on flux due to the
uncertainty in deriving an accurate value for the sky opacity $\tau$,
at a nominal airmass of 1.3. For observations at other airmasses ($A$)
the $y$ axis will scale as $\approx A/1.3$. The ordinate is a strong
function of $\tau$ and places a limit on the sky conditions which are
usable for 450$\mu$m observations when accurate calibration is
required. The crosses represent 450$\mu$m $\tau$ values, the stars are
850$\mu$m values. The scatter and relative uncertainty at 850$\mu$m
are much lower than at 450$\mu$m. The fitted line has the form
$y=-0.58+5.24\tau-6.97\tau^2+5.56\tau^3$.}
\end{figure}

The flux of an object at the top of the atmosphere is related to that
received by the telescope by the following relationship
\[
F_{obs,\lambda}=F_{a,\lambda}e^{-A\tau_{\lambda}}
\]
where $F_{obs,\lambda}$ is the observed flux at the telescope,
$F_{a,\lambda}$ is the flux at the top of the atmosphere before
attenuation, $A$ is the airmass of the source and $\tau_{\lambda}$ is
the zenith sky opacity. The opacity is wavelength dependent, being
greater at shorter wavelengths and is a strong function of atmospheric
water vapour content. Since $\tau$ is directly related to the
transmission and thus to the relative flux observed, it is important
to know its value as accurately as possible and to be measuring it
often enough to keep track of any variability. 
At constant airmass, a change in $\tau$ will produce the following
change in the flux:
\begin{equation}
\frac{F_1}{F_2}=\frac{e^{-A\tau_1}}{e^{-A\tau_2}}=e^{A\Delta\tau}
\label{fluxtauE}
\end{equation}
where $\Delta\tau$ is defined as $\tau_2-\tau_1$, and therefore
$F_1>F_2$. Expressed as a fractional change in flux, $\Delta
F/F=(F_1-F_2)/F_1$, this can be written as:
\begin{equation}
\frac{\Delta F}{F} = 1-e^{-A\Delta\tau}  \label{fracfluxE}
\end{equation}
We can make use of Eqn.\ref{fracfluxE} to estimate the overall
uncertainty in flux due to changes in $\tau$ in the following way: An
average value of $\tau$ is calculated from the four skydip fits
(varying the load temperatures) and the relationships with
$\tau_{850}$ and $\tau_{\rm{cso}}$ if available. The standard
deviation of these 5 or 6 estimates of $\tau$ can be calculated, as
well as the standard error on the mean value
($\sigma_{\tau_{\rm{av}}}$). This $\sigma_{\tau_{\rm{av}}}$ can then
be converted to a fractional error in the object flux using
Eqn.\ref{fracfluxE}, where
$\Delta\tau=\sigma_{\tau_{\rm{av}}}$. Fig.~\ref{tauF} shows $\Delta
F/F$ plotted against $\tau$, using the 450$\mu$m skydips at high
$\tau$ values and 850$\mu$m skydips for the lower $\tau$ values. The
uncertainty on the flux is a strong function of $\tau$. On nights with
high opacity the uncertainty in the fluxes is $\sim 7-15$\%, and for
$\tau \geq 1.7$ the inherent uncertainty on the flux due to the
extinction correction makes observing a pointless exercise if accurate
fluxes are required. This cannot be overcome by making more frequent
skydips (which is a separate issue), each point in Fig.~\ref{tauF} is
for a single point in time. Fig.~\ref{tauF} therefore gives a {\em
lower limit\/} to the extinction correction uncertainty. The problem
of the sampling rate of $\tau$, i.e. how $\tau$ varies between
skydips, will be discussed later. In the era before skydips, the accepted
practice was to observe the planetary calibrator at the same airmass as
the target object. This helps to reduce the effects of uncertainties in
$\tau$ and would still be of benefit to observers using skydips
today. One other point to note on observing techniques is that it is
not necessarily wise to focus the telescope only immediately before
planet measurements, as this means that the only maps in focus are
those of the calibrators, not the targets. This will enhance any
calibration problems from opacity correction and changes in the gain
(see next section). 

\renewcommand{\baselinestretch}{1.0}

\subsubsection{Changes in the gain}

The method used here to calibrate the fluxes relies on measuring the
calibration object in the same aperture as the source. The variation
of the calibration factor over time therefore sets an upper limit to
the accuracy of the 450$\mu$m galaxy fluxes. To asses this, all
of the 450$\mu$m calibration maps (13 maps from 11 nights) and a large
number of 850$\mu$m calibration maps (24 maps from 20 nights) were
taken and the signal (in Volts) measured in different sized apertures
(accounting for all the aperture sizes used to measure object
fluxes). The `aperture calibration factor',
$\rm{ACF}=S_{cal}(\rm{V})/S_{cal}(\rm{Jy})$, was then calculated for
each aperture on every map. The average ACF for each sized aperture at 450
and 850$\mu$m was taken and the standard deviation of all of the
measurements used to give the uncertainty on that value. Both the
average ACF and the uncertainties are listed in Table~\ref{gainT} and it
can be seen that the uncertainty is largest for the smallest
apertures, but plateaus out at $\sim 9-10$ percent at 450$\mu$m and 6
percent at 850$\mu$m. The last row (denoted $C_{\lambda}$) lists the
mean `standard' flux conversion factor (known as the FCF or gain) for
the calibration maps. This is what is generally used to calibrate
SCUBA maps, when the desired units are Jy
beam$^{-1}$. Table~\ref{gainT} shows that the values of $C_{\lambda}$
are far more variable than the aperture calibration factors (ACF) used
here to convert Volts to Jy. This is because measuring $C_{\lambda}$
uses the peak flux, which will be strongly affected by changes in beam
shape due to sky noise, pointing drifts, chop throw and the dish shape
(most critical at 450$\mu$m). Anything which broadens the beam will
remove flux from the peak and place it further away, making
$C_{\lambda}$ higher. The inherent surface accuracy and thermal
fluctuations in the shape of the dish cause large variations in
$C_{\lambda}$ at 450$\mu$m (the changes are still present, but smaller
at 850$\mu$m); thus the gain at 450$\mu$m often does not stabilise
until after midnight, and can vary by more than a factor of two in
the course of one night.

The results in Table~\ref{gainT} assume that long term changes in the
instrumental response and manual alterations to the dish shape (via
holography), are smaller than the short term changes in the beam
pattern due to thermal fluctuations and other effects. The validity of
this assumption was investigated by comparing observations made within
$\sim 1$ month of each other to see if the variation was any less. At
$450\mu$m this was not the case and so the values from
Table~\ref{gainT} represent the basic uncertainty in the ACF. However,
at $850\mu$m, a significant reduction was found when using
observations taken closer together (Table~\ref{850gainT}). The changes
in $C_{\lambda}$ over the 14 month period tie in with known
improvements/deteriorations in the dish surface. Therefore, at
$850\mu$m the values from Table~\ref{850gainT} should be used to
estimate the uncertainty in the ACF.

The ACF method presented here provides a more robust way to calibrate
fluxes, which is consistent to within 10 percent at 450$\mu$m and 2.6
percent at 850$\mu$m (in a typical aperture), over a time period of
$\sim 1$ month at 850$\mu$m and much longer at $450\mu$m, and in a
wide range of conditions.

\renewcommand{\baselinestretch}{1.0}
\begin{table*}
\centering
\caption{\label{gainT}Relative uncertainties in gains and calibration
factors used in flux measurement}
\begin{tabular}{cccccccc}
\hline
\\[-2ex]
\multicolumn{1}{c|}{Aperture}&\multicolumn{3}{c|}{850$\mu$m
(24)}&\multicolumn{3}{c|}{450$\mu$m (10)}&\multicolumn{1}{c}{Notes}\\
\multicolumn{1}{c|}{}&\multicolumn{1}{c}{mean}&\multicolumn{1}{c}{$\sigma$}&\multicolumn{1}{c|}{\% 
$\sigma$}&\multicolumn{1}{c}{mean}&\multicolumn{1}{c}{$\sigma$}&\multicolumn{1}{c|}{\% 
$\sigma$}&\multicolumn{1}{c}{}\\
\\[-2ex]
\hline
\\[-2ex]
15$^{\prime\prime}$ & 2.229 & 0.167 & 7.5 & 11.13 & 1.37 & 12 & Aperture\\
30$^{\prime\prime}$ & 1.134 & 0.080 & 7.0 & 7.20 & 0.63 & 9 & calibration\\
45$^{\prime\prime}$ & 1.013 & 0.060 & 6.0 & 5.79 & 0.49 & 9 & method\\
60$^{\prime\prime}$ & 0.942 & 0.052 & 5.5 & 5.12 & 0.50 & 10 &\\
90$^{\prime\prime}$ & 0.879 & 0.050 & 5.6 & 4.75 & 0.50 & 10 &\\
\\[-2.5ex]
\\[-2.5ex]
$C_{\lambda}$ & 278 & 22 & 8.0 & 815 & 144 & 18 & `Standard' peak gain
or FCF\\
\\[-2.5ex]
\hline
\end{tabular}
\flushleft
\small{Mean, standard deviation and \%
variation in the `aperture calibration factors' and the `standard'
peak calibration factor for both wavelengths over the entire course of
the observations. Numbers in parentheses are the numbers of maps used
to determine these figures. The method of calibration is in the final
column. At $450\mu$m, the variation here is no greater than that over
a shorter time period but for $850\mu$m it is an overestimate, see
Table~\ref{850gainT}.}
\end{table*}

\begin{table*}
\centering
\caption{\label{850gainT} Variation in gains and calibration factors
at 850$\mu$m over shorter time scales than Table~\ref{gainT}.}
\begin{tabular}{cc|cc|ccccc}
\\[-2ex]
\hline
\\[-2ex]
\multicolumn{1}{c}{Dates}&\multicolumn{1}{c|}{$\rm{N_{maps}}$}&\multicolumn{2}{c|}{$C_{\lambda}$}&\multicolumn{5}{c}{\% 
change in ACF}\\
\multicolumn{2}{c|}{}&\multicolumn{1}{c}{Mean}&\multicolumn{1}{c|}{\% $\sigma$}&\multicolumn{1}{c}{$15^{\prime\prime}$}&\multicolumn{1}{c}{$30^{\prime\prime}$}&\multicolumn{1}{c}{$45^{\prime\prime}$}&\multicolumn{1}{c}{$60^{\prime\prime}$}&\multicolumn{1}{c}{$90^{\prime\prime}$}\\
\\[-2ex]
\hline
\\[-2ex]
July/Aug 97 & 4 & 266 & 3.0 & 2.5 & 2.3 & 1.4 & 0.4 & 0.5\\
Dec/Jan 97 & 4 & 283 & 4.0 & 5.0 & 4.4 & 4.4 & 3.6 & 4.1\\
March 98 & 9 & 259 & 4.3 & 3.1 & 2.1 & 1.8 & 1.8 & 2.3\\
June/July 98 & 5 & 293 & 6.0 & 4.5 & 3.8 & 2.9 & 2.3 & 2.4\\
Sept 98 & 2 & 294 & 5.7 & 5.2 & 4.1 & 2.8 & 1.7 & 1.3\\
\\[-2ex]
\hline
\\[-2ex]
\multicolumn{3}{c}{Average}&\multicolumn{1}{c}{4.6}&\multicolumn{1}{c}{4.0}&\multicolumn{1}{c}{3.3}&\multicolumn{1}{c}{2.6}&\multicolumn{1}{c}{2.0}&\multicolumn{1}{c}{2.1}\\
\\[-2ex]
\hline
\end{tabular}
\flushleft
\small{850$\mu$m \% variation in $C_{\lambda}$
and ACF for time periods of $\sim 1$ month. The mean $C_{\lambda}$ can
be seen to change on these time scales due to deliberate alterations
in the dish shape. As a result of this, the variations within these
time frames are smaller than those seen in Table~\ref{gainT} and thus
is it the values from this Table which should be used in estimating
$\sigma_{\rm{gain}}$ at $850\mu$m.}
\end{table*}

The total calibration uncertainty at 450$\mu$m can be estimated as
follows:

\begin{equation}
\sigma_{\rm{cal}}=(\sigma_{\tau}^2 +
\sigma_{\rm{gain}}^2+\sigma_{\rm{vary}}^2 + \sigma_{\rm{abs}}^2)^{1/2}  \label{calerrE}
\end{equation}
where $\sigma_{\tau}$ is the uncertainty in $\tau$ from skydip fitting
(on average 7 per cent for these observations, see Fig.~\ref{tauF}),
$\sigma_{\rm{gain}}$ is the uncertainty in the ACF factor (on average
9 per cent, see Table~\ref{gainT}), $\sigma_{\rm{abs}}$ is the
absolute uncertainty in the brightness temperatures of the planets
which are used for calibration (estimated to be 5 per cent (SUN/213))
and $\sigma_{\rm{vary}}$ represents the uncertainty in how the sky
opacity has changed between skydips. This is not really calculable, as
it will be a function of skydip frequency and atmospheric
instability. However, the calibrators themselves have had their fluxes
corrected for extinction, and so $\sigma_{\rm{vary}}$ must be part of
$\sigma_{\rm{gain}}$. We estimate an upper limit on $\sigma_{\rm{vary}}$ as
$\sqrt{\sigma_{\rm{gain}}^2-\sigma_{\tau}^2}$ which is $\approx 2-4$
percent. Including all terms, the total calibration uncertainty is
$\sim 10$ percent at 850$\mu$m and 15 percent at 450$\mu$m.

\section{Results}
Table~\ref{450fluxT} lists the 450$\mu$m fluxes which we analysed and
which were measured in the way described in Section~\ref{obsS} (this
includes the archival data). Fluxes from the literature for the SLUGS
galaxies are listed in Table~\ref{subfluxT} and are from a variety
telescopes (ISO, JCMT, IRAM, OVRO). Other fluxes for Arp220 are listed
separately in Table~\ref{a220T}. There are complications when
making comparisons between fluxes measured at different wavelengths if
there are significant beam size differences between them. We did not
include fluxes from the literature if the beam sizes were smaller than
the extent of the submm emission.

\subsection{\label{450corrS}Corrections applied to fluxes}
\subsubsection{SCUBA maps}

We measured the 450$\mu$m emission in the same sized aperture used for
the 850$\mu$m observations, which was chosen to maximise the
signal-to-noise (S/N) at the longer wavelength. However, because of
the lower S/N at the shorter wavelengths, the 450$\mu$m emission is
sometimes only clearly detected in the central region of the aperture,
and systematic effects, such as those caused by poor sky subtraction
are more of a concern. We investigated this by convolving the 450 and
850$\mu$m maps to the same resolution and taking a small aperture
around the bright central region on both maps. The ratio of the
450$\mu$m and 850$\mu$m fluxes in this region was compared to that
measured using the larger aperture. If the ratio for the larger
aperture was significantly lower than the central ratio, the total
450$\mu$m flux would be corrected by a factor which would reproduce
the central ratio across the whole galaxy. In doing this, it is
implicitly assumed that the true flux ratio is constant across the
galaxy. Observations of the large, well-resolved galaxies NGC 891 and
NGC 7331 (Alton et al. 1998; Bianchi et al. 1998; Israel et al. 1999;
Alton et al. 2001) and of the interacting pair VV114 (Frayer et
al. 1999), and of the Milky Way (Sodroski et al. 1997) suggest that
the actual change in $S_{450}/S_{850}$ due to dust temperature
gradients would be only $\sim 14$ percent or less. In practise, such a
correction was only necessary for four objects, two of which (NGC 5962
and NGC 7541) were large and extended over most of the array. The
other two (UGC 2369 and UGC 2403) were observed in quite noisy sky
conditions on the same night. The corrections were not very large,
ranging from 10--34 per cent and they do not affect the
conclusions. Fluxes in Table~\ref{450fluxT} have been corrected for
this effect, and any corrections used are also listed.

\renewcommand{\baselinestretch}{1.0}
\begin{table*}
\centering
\caption{\label{450fluxT}Flux densities for the SLUGS galaxies}
\begin{tabular}{lccccccccccc}
\hline
\\[-2.5ex]
\multicolumn{1}{c}{(1)}&\multicolumn{1}{c}{(2)}&\multicolumn{1}{c}{(3)}&\multicolumn{1}{c}{(4)}&\multicolumn{1}{c}{(5)}&\multicolumn{1}{c}{(6)}&\multicolumn{1}{c}{(7)}&\multicolumn{1}{c}{(8)}&\multicolumn{1}{c}{(9)}&\multicolumn{1}{c}{(10)}&\multicolumn{1}{c}{(11)}&\multicolumn{1}{c}{(12)}\\
\\[-1ex]
\multicolumn{1}{c}{Name}&\multicolumn{1}{c}{R.A.}&\multicolumn{1}{c}{Decl.}&\multicolumn{1}{c}{$cz$}&\multicolumn{1}{c}{$S_{60}$}&\multicolumn{1}{c}{$S_{100}$}&\multicolumn{1}{c}{$S_{450}$}&\multicolumn{1}{c}{$\sigma_{450}$}&\multicolumn{1}{c}{$S_{850}$}&\multicolumn{1}{c}{$\sigma_{850}$}&\multicolumn{1}{c}{$f_{cor}$}&\multicolumn{1}{c}{$\frac{S_{450}}{S_{850}}$}\\
\multicolumn{1}{c}{}&\multicolumn{1}{c}{(J2000)}&\multicolumn{1}{c}{(J2000)}&\multicolumn{1}{c}{(km
s$^{-1}$)}&\multicolumn{1}{c}{(Jy)}&\multicolumn{1}{c}{(Jy)}&\multicolumn{1}{c}{(mJy)}&\multicolumn{1}{c}{(mJy)}&\multicolumn{1}{c}{(mJy)}&\multicolumn{1}{c}{(mJy)}&\multicolumn{1}{c}{}&\multicolumn{1}{c}{}\\
\\[-2.5ex]
\hline
\\[-2.5ex]
UGC 903 & 1 21 47.9 & $+$17 35 34 & 2518 & 7.91 & 14.58 & 1500 & 315 & 
178 & 26 & .. & 8.43\\
NGC 958 & 2 30 42.8 & $-$02 56 23 & 5738 & 5.90 & 14.99 & 2251 & 428 & 
262 & 34 & .. & 8.60\\ 
UGC 2369 & 2 54 01.8 & $+$14 58 14 & 9400 & 7.68 & 11.10 & 523 & 120 & 
72 & 13 & 1.11 & 7.26\\
UGC 2403 & 2 55 57.2 & $+$00 41 33 & 4161 & 7.51 & 11.77 & 1010 & 202
& 111 & 18 & 1.26 & 9.10\\
NGC 1614$^A$ & 4 34 00.0 & $-$08 34 45 & 4778 & 33.12 & 36.19 & 981 & 167
& 140 & 20 & .. & 7.45\\ 
NGC 1667 & 4 48 37.2 & $-$06 19 12 & 4547 & 6.24 & 16.54 & 1183 & 272
& 163 & 22 & .. & 7.26\\
NGC 2856 & 9 24 16.2 & $+$49 14 58 & 2638 & 6.15 & 10.28 & 993 & 268 & 
89 & 16 & .. & 11.16\\
NGC 2990 & 9 46 17.2 & $+$05 42 33 & 3088 & 5.49 & 10.16 & 1275 & 332
& 110 & 19 & .. & 11.60\\
UGC 5376 & 10 00 26.8 & $+$03 22 26 & 2050 & 5.94 & 11.49 & 1258 & 415 
& 148 & 23 & .. & 8.50\\
ARP 148 & 11 03 54.0 & $+$40 50 59 & 10350 & 6.95 & 10.99 & 646 &
156 & 92 & 20 & .. & 7.02\\
MCG+00-29-023 & 11 21 12.2 & $-$02 59 03 & 7464 & 5.40 & 8.87 & 571 &
166 & 84 & 13 & .. & 6.80\\
ZW 247.020 & 14 19 43.3 & $+$49 14 12 & 7666 & 5.91 & 8.25 & 284 & 111 
& 36 & 8 & .. & 7.90\\
1 Zw 107$^A$ & 15 18 06.1 & $+$42 44 45 & 11946 & 9.15 & 10.04 & 423 & 93
& 60 & 14 & .. & 7.05 \\
IR 1525+36 & 15 26 59.4 & $+$35 58 37 & 16009 & 7.20 & 5.78 & 252 & 70 
& 33 & 8 & .. & 7.64 \\
ARP 220 & 15 34 57.2 & $+$23 30 11 & 5452 & 103.33 & 113.95 & 6286 &
786 & 832 & 86 & .. & 7.56\\
NGC 5962 & 15 36 32.0 & $+$16 36 22 & 1963 & 8.99 & 20.79 & 1959 & 372 
& 317 & 37 & 1.12 & 6.18\\
NGC 6052 & 16 05 13.0 & $+$20 32 34 & 4712 & 6.46 & 10.18 & 721 & 230
& 95 & 15 & .. & 7.59\\
NGC 6181 & 16 32 21.2 & $+$19 49 30 & 2379 & 9.35 & 21.00 & 1470 & 456 
& 228 & 37 & .. & 6.45\\
NGC 7541 & 23 14 43.4 & $+$04 32 04 & 2665 & 20.59 & 40.63 & 2639 &
686 & 427 & 60 & 1.34 & 6.18\\
\hline
\end{tabular}
\flushleft
\small (1) Most commonly used name taken
from the IRAS BGS (Soifer et al. 1989). (2) Right ascension J2000
epoch.  (3) Declination J2000 epoch. (4) Recession velocity taken from
NED. [The NASA/IPAC Extragalactic Database (NED) is operated by the
Jet Propulsion Laboratory, California Institute of Technology, under
contract with the National Aeronautics and Space Administration.]  (5)
60$\mu$m flux from BGS. (6) 100$\mu$m flux from BGS.  (7) 450$\mu$m
flux (this work). (8) error on 450$\mu$m flux, calculated in the
manner described in Section~\ref{obsS} and inclusive of a 15\%
calibration uncertainty.  (9) 850$\mu$m flux (Paper I). N.B. The
values for NGC 1614 and IR 1525+36 have been revised from those listed
in Paper I in light of new data.  (10) error on 850$\mu$m flux (Paper
I, includes a 10\% calibration uncertainty). (11) Correction factor
applied to $S_{450}$ if a decline in $S_{450}/S_{850}$ was found
(Section~\ref{450corrS}); corrected values are used in Cols 5,7,12 and
later in this paper. (12) Ratio of 450$\mu$m to 850$\mu$m fluxes. $^A$
indicates that the data was taken from the JCMT archive..
\end{table*}

\renewcommand{\baselinestretch}{1.0}
\begin{table*}
\centering
\caption{\label{subfluxT} Other submillimetre fluxes for the IRAS
sample from the literature.}
\begin{tabular}{lcccccccccc}
\\[-2ex] 
\hline
\\[-2.5ex]
\multicolumn{1}{c}{(1)}&\multicolumn{1}{c}{(2)}&\multicolumn{1}{c}{(3)}&\multicolumn{1}{c}{(4)}&\multicolumn{1}{c}{(5)}&\multicolumn{1}{c}{(6)}&\multicolumn{1}{c}{(7)}&\multicolumn{1}{c}{(8)}&\multicolumn{1}{c}{(9)}&\multicolumn{1}{c}{(10)}&\multicolumn{1}{c}{(11)}\\
\multicolumn{1}{c}{Name}&\multicolumn{1}{c}{R.A.}&\multicolumn{1}{c}{Decl.}&\multicolumn{1}{c}{$cz$}&\multicolumn{1}{c}{$S_{60}$}&\multicolumn{1}{c}{$S_{100}$}&\multicolumn{1}{c}{$S_{850}$}&\multicolumn{1}{c}{$\sigma_{850}$}&\multicolumn{1}{c}{$\lambda$}&\multicolumn{1}{c}{$S_{\lambda}$}&\multicolumn{1}{c}{Refs.}\\
\multicolumn{1}{c}{}&\multicolumn{1}{c}{(J2000)}&\multicolumn{1}{c}{(J2000)}&\multicolumn{1}{c}{(km
s$^{-1}$)}&\multicolumn{1}{c}{(Jy)}&\multicolumn{1}{c}{(Jy)}&\multicolumn{1}{c}{(mJy)}&\multicolumn{1}{c}{(mJy)}&\multicolumn{1}{c}{($\mu$m)}&\multicolumn{1}{c}{(Jy)}&\multicolumn{1}{c}{}\\
\\[-2.5ex]
\hline
\\[-2.5ex]
NGC 520 &  1 24 34.9 & $+$03 47 31 & 2281 & 31.55 & 46.56 & 325 & 50 & 
1200 & 0.076 & a\\
UGC 2982 & 4 12 22.5 & $+$05 32 51 & 5305 & 8.70 & 17.32 & 176 & 34 & 
180, 190, 1250 & 9.9, 9.2, 0.047 & b,c\\
NGC 2623 & 8 38 24.1 & $+$25 45 16 & 5535 & 25.72 & 27.36 & 91 & 14 & 350, 750 & 2.225, 0.170 & d\\
NGC 3110 & 10 04 02.0 & $-$06 28 31 & 5048 & 11.68 & 23.16 & 188 & 28
& 1250 & 0.048 & c\\
IR 1017+08 & 10 20 00.2 & $+$08 13 34 & 14390 & 6.08 & 5.97 & 36 & 6 & 
350, 750 & 0.450, 0.049 & d\\
IR 1056+24 &  10 59 18.2 & $+$24 32 34 & 12912 & 12.53 & 16.06 & 61 & 13
& 350, 450, 750 & 1.24, 0.533, 0.085 & d\\
IR 1211+03 &12 13 46.1 & $+$02 48 40 & 21703 & 8.39 & 9.10 & 49 & 10
& 450 & 0.429 &  d,e\\
NGC 4418$^{\ast}$ & 12 26 54.7 & $-$00 52 39 & 2179 & 42.32 & 30.76 & 255 & 37
& 450, 800, 1100 & 1.474, 0.300, 0.093 &  f\\
UGC 8387 & 13 20 35.3 & $+$34 08 22 & 7000 & 13.69 & 24.90 & 113 & 15
& 350, 750 & 1.946, 0.196 & d\\
Zw 049.057 & 15 13 13.1 & $+$07 13 31 & 3927 & 21.06 & 29.88 & 200 &
27 & 1250 & 0.047 &  c\\
NGC 7592 & 23 18 22.1 & $-$04 24 58 & 7350 & 8.02 & 10.50 & 108 & 19 & 
80, 180, 200 & 11.2, 5.8, 4.6 &  b\\
NGC 7679 & 23 28 46.7 & $+$03 30 41 & 5138 & 7.28 & 10.65 & 93 & 15 & 80, 180, 200 & 9.4, 5.8, 4.6 &  b\\
NGC 7714 & 23 36 14.1 & $+$02 09 18 & 2798 & 10.52 & 11.66 & 72 & 13 & 80, 180, 200 & 11.0, 4.9, 4.3 & b\\
\\[-2.5ex]
\hline
\end{tabular}
\flushleft
\small (1) Most commonly used name taken
from the IRAS BGS (Soifer et al. 1989). (2) Right ascension J2000 epoch.
(3) Declination J2000 epoch. (4) Recession velocity taken from NED.
(5) 60$\mu$m flux from BGS. (6) 100$\mu$m flux from BGS. (7) 850$\mu$m
flux (Paper I). (8) error on 850$\mu$m flux (Paper I, includes a 10\%
calibration uncertainty). (9) Wavelengths of
measurements (microns). (10) Fluxes at each $\lambda$ in Jy. (11)
References for fluxes: a) Braine \& Dumke 1998; b) Siebenmorgen et
al. 1999; c) Carico et al. 1992; d) Fox 2000; e) Rigopoulou et
al. 1996; f) Roche \& Chandler 1993; g) Klaas et al. 1997; h) Downes
\& Solomon 1998; i) Woody et al. 1989\\
$^{\ast}$ Fluxes for NGC 4418 have been beam corrected using the SCUBA
850$\mu$m map (Section~\ref{450corrS}). Corrections were small (10\%
increase at 450$\mu$m, 25\% at 800$\mu$m and 9\% at 1100).
\end{table*}

\renewcommand{\baselinestretch}{1.0}
\begin{table}
\centering
\small
\caption{\label{a220T} Other FIR/submm fluxes for Arp220.}
\begin{tabular}{ccc}
\\[-2ex] 
\hline
\\[-2.5ex]
\multicolumn{1}{c}{Wavelength}&\multicolumn{1}{c}{Flux}&\multicolumn{1}{c}{Ref.}\\
\multicolumn{1}{c}{($\mu$m)}&\multicolumn{1}{c}{(Jy)}&\multicolumn{1}{c}{}\\
\\[-2.5ex]
\hline
\\[-2.5ex]
60 & 109.4 & g\\
65 & 114.4 & g\\
90 & 95.2  & g\\
100 & 105.3 & g\\
120 & 73.4  & g\\
150 & 58.5  & g\\
170 & 47.9  & g\\
180 & 37.5  & g\\
200 & 34.8  & g\\
350 & 11.7  & e\\
450 & 6.286 & *\\
850 & 0.832 & *\\
1100 & 0.350 & e\\
1250 & 0.226 & c\\
1300 & 0.175 & h\\
1400 & 0.140 & i\\  
\\[-2.5ex]
\hline
\end{tabular}
\flushleft
\small References as for
Table~\ref{subfluxT}, $^*$ submm fluxes from this work. ISO fluxes (g) are colour corrected as described in that paper.
\end{table}

\subsubsection{Literature fluxes}
Due to the uncertainty involved in making beam corrections, only
literature fluxes which were obtained with single-element detectors
with beams larger than the area of SCUBA emission (i.e. long
wavelength ISO data, observations of very small
sources) or which come from maps were used. However, a small
correction was applied to the fluxes for NGC 4418, which was observed
at 450, 800 and 1100$\mu$m by Roche \& Chandler (1993) using UKT14 on
the JCMT. This source is very compact with almost all of the flux in
one SCUBA beam. The SCUBA map was used to find the 850$\mu$m flux in
the equivalent region to the UKT14 beam at each wavelength. In this
way correction factors of 1.10, 1.25 and 1.09 were derived for the
450, 800 and 1100$\mu$m fluxes respectively. This is a small
difference and has very little effect on the derived parameters in
later discussions.


\subsection{\label{2sedS}Two-component Spectral Energy Distributions.}

The emission at a particular frequency will now be represented
as the sum of two modified Planck functions, each with a
different characteristic temperature. In reality, the dust is probably
at a range of temperatures reflecting the localised heating sources in
the ISM, but the approximation of characteristic temperatures for the
warmer and cooler dust is not a bad one. For the optically thin
regime, this can be expressed as:
\begin{equation}
S_{\nu}=N_w\times\nu^{\beta}{\rm{B}}(\nu,\,T_w) + \label{2compSEDE}
N_c\times\nu^{\beta}{\rm{B}}(\nu,\,T_c) 
\end{equation}
where $N_w$ and $N_c$ represent the relative masses in the warm and
cold components, $T_w$ and $T_c$ are the temperatures,
${\rm{B}}(\nu,\,T)$ is the Planck function for each component, and
$\beta$ is the dust emissivity index (assumed to be the same for
each). This model was fitted to the fluxes and the
parameters which produced the minimum $\chi^2$ were found. Initially
the value of $\beta$ was allowed to vary between $1\rightarrow 2$;
$T_w$ was constrained by the IRAS 25$\mu$m flux, which was not allowed
to be exceeded in a fit (N.B. a fit was not made {\em to\/} this point
as this would require an additional hot component); $T_c$ was allowed
to take any value lower than $T_w$. For sources with only 4 fluxes
(not including 25$\mu$m), there are not enough data points to provide
a well-constrained fit, consequently the values of $\chi^2_{\rm{min}}$
are unrealistically low and a large range of parameters will provide
very acceptable fits. The parameters producing the best fits are in
Table~\ref{sed1T} along with the original isothermal SED parameters
from Paper I -- derived from fits to the 60, 100 and 850$\mu$m points
only. Uncertainties in the parameters from the two-component fits are
also given and these were produced using the bootstrap method, as
described in Paper I.

\renewcommand{\baselinestretch}{1.0}
\begin{figure}
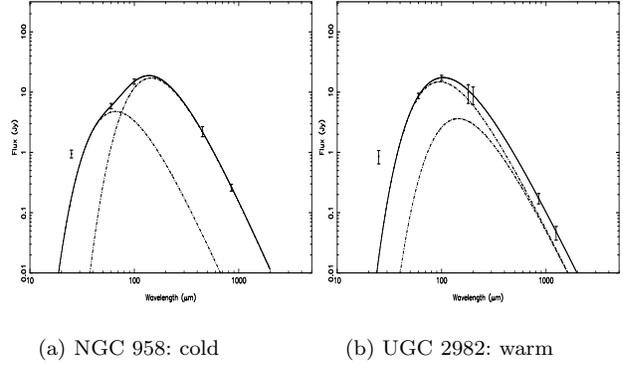

\centerline{
 \subfigure[NGC 958: cold]{\psfig{file=Fig2a.ps,width=4cm,height=4cm,angle=-90}}\\
 \subfigure[UGC 2982: warm]{\psfig{file=Fig2b.ps,width=4cm,height=4cm,angle=-90}}\\}
\caption{\label{coldsedF} a) SED of galaxy with a prominent cold
component, not compatible with a single temperature SED. b) SED of
galaxy with dominant warm component, any cold dust is very difficult
to see in the SED (if it is there at all). The solid lines indicate
the composite two-component SED (parameters listed in
Table~\ref{sed1T}). The dot-dashed lines represent the warm and cold components.}
\end{figure}

\renewcommand{\baselinestretch}{1.0}
\begin{figure}
\centerline{\psfig{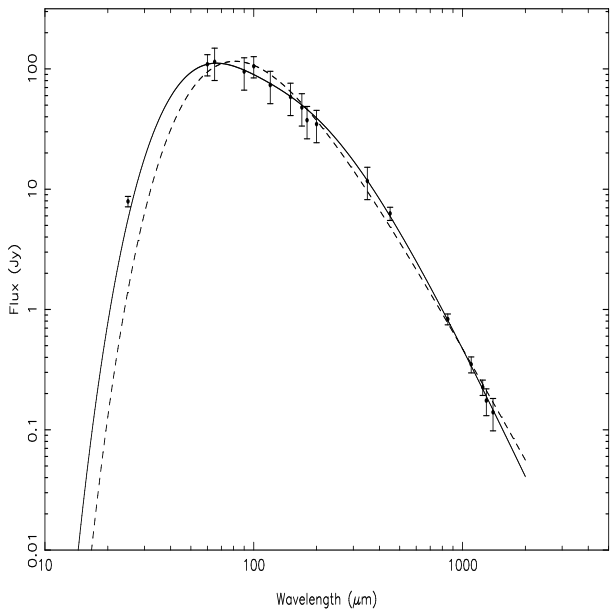}}
\caption{\label{A220F} Best fitting SEDs for Arp 220 when $\beta$ is a
free parameter. Although there are more than enough data points
determine the free parameters $T_w$, $T_c$, $\beta$ and $N_c/N_w$ it
is still possible to fit SEDs with either single temperatures (dashed)
and lower $\beta$, or two temperatures and steeper $\beta$ (solid). The
fundamental problem in confirming the existence of colder components
is not simply one of a lack of data points, but rather a degeneracy in
the parameters. The parameters for this fit are listed in
Table~\ref{sed1T}.}
\end{figure}

\renewcommand{\baselinestretch}{1.0}
\begin{table*}
\centering
\caption{\label{sed1T} SED parameters for isothermal and free $\beta$ two-component fits.}
\begin{tabular}{lccccccccccc}
\\[-2ex] 
\hline
\\[-2.5ex]
\multicolumn{1}{c}{(1)}&\multicolumn{1}{c}{(2)}&\multicolumn{1}{c}{(3)}&\multicolumn{1}{c}{(4)}&\multicolumn{1}{c}{(5)}&\multicolumn{1}{c}{(6)}&\multicolumn{1}{c}{(7)}&\multicolumn{1}{c}{(8)}&\multicolumn{1}{c}{(9)}&\multicolumn{1}{c}{(10)}&\multicolumn{1}{c}{(11)}&\multicolumn{1}{c}{(12)}\\
\multicolumn{1}{c}{Name}&\multicolumn{1}{c}{$\beta_s$}&\multicolumn{1}{c}{$T_{\rm{d}}$}&\multicolumn{1}{c}{$\beta_2$}&\multicolumn{1}{c}{$\sigma_{\beta_2}$}&\multicolumn{1}{c}{$T_w$}&\multicolumn{1}{c}{$\sigma_{T_w}$}&\multicolumn{1}{c}{$T_c$}&\multicolumn{1}{c}{$\sigma_{T_c}$}&\multicolumn{1}{c}{$\frac{N_c}{N_w}$}&\multicolumn{1}{c}{$\sigma_{N}$}&\multicolumn{1}{c}{$N_{\lambda}$}\\
\multicolumn{1}{c}{}&\multicolumn{1}{c}{}&\multicolumn{1}{c}{(K)}&\multicolumn{1}{c}{}&\multicolumn{1}{c}{}&\multicolumn{1}{c}{(K)}&\multicolumn{1}{c}{(K)}&\multicolumn{1}{c}{(K)}&\multicolumn{1}{c}{(K)}&\multicolumn{1}{c}{}&\multicolumn{1}{c}{}&\multicolumn{1}{c}{}\\
\\[-2.5ex]
\hline
\\[-2.5ex]
UGC 903 & 1.2 & 34.4 & 2.0 & (0.36) & 45 & (5.1) & 21 & (4.2) & 91 &
(52) & 4\\ NGC 958 & 1.2 & 30.8 & 2.0 & (0.25) & 44 & (9.1) & 20 &
(2.9) & 186 & (341) & 4\\ UGC 2369 & 1.4 & 36.2 & 1.5 & (0.35) & 41 &
(13.3) & 32 & (6.0) & 3.3 & (101) & 4\\ UGC 2403 & 1.2 & 36.8 & 2.0 &
(0.30) & 50 & (6.5) & 22 & (4.8) & 100 & (65) & 4\\ NGC 1614 & 1.6 &
39.2 & 1.6 & (0.23) & 40 & (9.6) & 31 & (9.1) & 0.06 & (20) & 4\\ NGC
1667 & 1.6 & 28.4 & 1.6 & (0.27) & 31 & (10.2) & 28 & (4.4) & 3.9 &
(136) & 4\\ NGC 2856 & 1.3 & 35.0 & 2.0 & (0.27) & 45 & (8.7) & 22 &
(4.5) & 58 & (134) & 4\\ NGC 2990 & 1.3 & 33.8 & 2.0 & (0.17) & 42 &
(8.9) & 21 & (2.6) & 58 & (133) & 4\\ UGC 5376 & 1.2 & 33.8 & 2.0 &
(0.42) & 44 & (7.4) & 21 & (5.5) & 87 & (116) & 4\\ Arp 148 & 1.3 &
35.6 & 1.5 & (0.35) & 47 & (7.6) & 30 & (6.5) & 15.5 & (31) & 4\\
MCG+00-29-023 & 1.3 & 35.0 & 1.5 & (0.32) & 40 & (9.3) & 27 & (5.8) &
4.8 & (50) & 4\\ Zw 247.020 & 1.7 & 34.4 & 1.7 & (0.24) & 36 &(14.0)&
28 &(5.4)& 0.4 &(97) & 4\\ 1 Zw 107 & 1.3 & 41.6 & 1.4 & (0.39) & 43 &
(10.5) & 27 & (8.3) & 0.9 & (30) & 4\\ IR 1525+36 & 1.1 & 53.0 & 1.7 &
(0.39) & 57 & (5.8) & 26 & (9.7) & 15.2 & (17) & 4\\ NGC 5962 & 1.2 &
32.0 & 1.4 & (0.28) & 32 & (9.8) & 21 & (7.3) & 1.0 & (77) & 4\\ NGC
6052 & 1.3 & 35.6 & 1.7 & (0.34) & 55 & (12.8) & 26 & (6.8) & 65 &
(65) & 4\\ NGC 6181 & 1.4 & 30.8 & 1.4 & (0.33) & 32 & (11.6) & 31 &
(5.3) & 1.0 & (165) & 4\\ NGC 7541 & 1.3 & 33.2 & 1.4 & (0.31) & 34 &
(9.6) & 27 & (4.8) & 0.6 & (63) & 4\\
\\[-2ex]
\hline
\\[-2ex]
NGC 520 & 1.4 & 36.2 & 2.0 & (0.24) & 50 & (6.4) & 24 & (4.4) & 61 & (42) & 4\\
UGC 2982 & 1.4& 32.0 & 1.6 & (0.27) & 33 & (10.7) &
22 & (6.7) & 1.6 & (872) & 6\\
NGC 2623 & 1.6 & 39.8 & 2.0 & (0.15) & 50 & (5.7) & 27 & (4.5)
& 18 & (17) & 5\\
NGC 3110 & 1.5 & 32.0 & 1.9 & (0.30) & 31 & (12.4) & 18 & (10.1) & 2.7 &
(83) & 4 \\
IR 1017+08 & 1.2 & 45.2 & 1.3 & (0.32) & 44 & (8.2) & 27 &
(13.3) & 0.3 & (18) & 5 \\
IR 1056+24 & 1.7 & 35.6 & 2.0 & (0.23) & 40 &
(12.1) & 26 & (6.6) & 5.6 & (67) & 6 \\
IR 1211+03 & 1.3 & 42.2 & 1.9 & (0.34) & 49 & (5.5) & 25 & (7.5) & 20 & (44) & 4 \\
NGC 4418 & 0.8 & 62.0 & 1.5 & (0.35) & 55
& (6.3) & 23 & (6.2) & 8.9 & (25) & 6 \\
UGC 8387 & 1.8 & 30.8 & 1.8 & (0.16) & 32 & (11.3) & 31 & (3.8)
& 0.7 & (1340) & 5 \\
Zw 049.057 & 1.4 & 36.2 & 2.0 & (0.33) & 44 & (5.0) & 23 & (7.0) & 28 & (35) & 4 \\
Arp 220 & 1.2 & 42.2 & 1.8 & (0.24) & 47 & (3.6) &
19 & (3.6) & 19 & (14) & 16\\
NGC 7592 & 1.1 & 39.8 & 1.5 & (0.33) & 39 & (11.0) & 20
& (11.1) & 3.5 & (76) & 6 \\
NGC 7679 & 1.2 & 38.0 & 1.5 & (0.29) & 48 & (11.8) & 29 & (6.5)
& 13.7 & (64) & 6 \\
NGC 7714 & 1.3 & 41.0 & 1.5 & (0.30) & 60 & (12.8) & 32 
& (7.9) & 19.0 & (171) & 6 \\
\\[-2.5ex]
\hline
\end{tabular}
\flushleft
\small (1) Galaxy name. (2) Emissivity index derived from an isothermal
fit to the 60, 100 and 850$\mu$m fluxes (Paper I). (3) Dust
temperature derived from an isothermal fit to the 60, 100 and
850$\mu$m fluxes (Paper I). (4--5) Best-fitting value of dust
emissivity index $\beta$ using a two-component SED and 1$\sigma$
uncertainty from bootstrap calculation. (6--9) Fitted warm and cold
component temperatures with 1$\sigma$ uncertainties. (10--11) The
ratio of cold dust mass to warm dust mass and 1$\sigma$
uncertainty. (12) Number of fluxes used in the two-component fit (not
including the 25$\mu$m point).\\ Galaxies below the line are those
with extra FIR/submm fluxes taken from the literature while above are
those for which we have our own data.
\end{table*}

The relative contribution of the cold component to the SEDs shows a
large variation (the best parameter to describe this would be
$N_c/N_w$). Fig.~\ref{coldsedF}a shows an example of a very `cold' SED
where the cold component is clearly visible and, once the 450$\mu$m
point is included, a single temperature fit is excluded. This is in
contrast to other objects (Fig.~\ref{coldsedF}b) where a cold
component can be included but its contribution is barely noticeable
and it is statistically indistinguishable from an isothermal SED. The
problem with indeterminate objects such as that in Fig.~\ref{coldsedF}b
is not simply due a limited number of data points. For example, the
SED of Arp 220 (Fig.~\ref{A220F}) has more than enough data points, yet
can still be fitted by either single temperature models with lower
$\beta$ or two-component models with higher $\beta$ (however
two-component models produce the {\em best\/} fits).

\begin{figure}
\centerline{\psfig{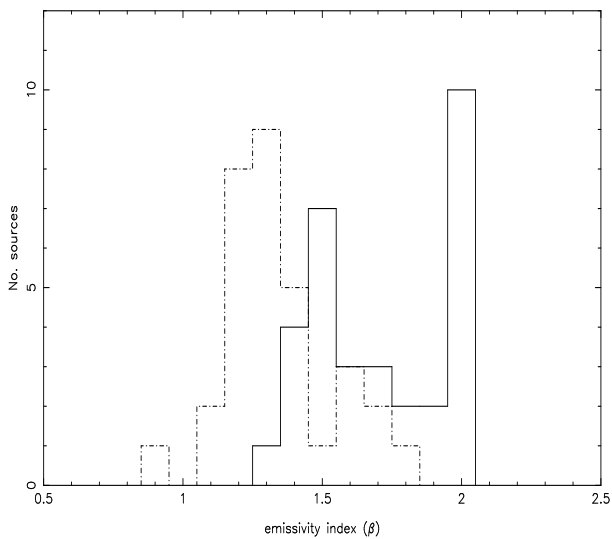}}
\caption[Distribution of $\beta$ for the 32
sources with other submm fluxes.]{\label{bhistx2F} Distribution of
$\beta$ for the 32 sources with submm fluxes at more than one
wavelength. The solid line is the 2 component $\beta$ distribution and
the dot-dash line represents the $\beta$ from the original single
temperature fits to the 60, 100 and 850$\mu$m fluxes.}
\end{figure}

\begin{figure}
\centerline{\psfig{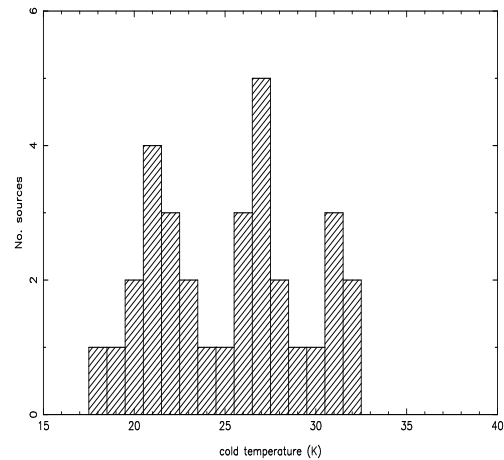}}
\caption[Distribution of cold and warm
temperatures.]{\label{temphistF} Distribution of cold temperatures
when $\beta$ is allowed to take on values between 1--2.}
\end{figure}

The distribution of $\beta$ for the sample is presented in
Fig.~\ref{bhistx2F}, where the original single temperature $\beta_s$ for
these galaxies are also shown. The two-component $\beta$ distribution
(solid) is shifted toward higher values than the single temperature
one (dot-dashed) and also seems to have two maxima, one at $\beta=2$
and another at $\beta=1.5$. The $T_c$ distribution
(Fig.~\ref{temphistF}) is also quite odd with 2--3 peaks at $\sim 22,
27$ and 31 K. The galaxies producing the $\beta\approx1.5$ peak seem
to be the same ones responsible for the higher $T_c$ peaks at 27--31
K. This bi-modal behaviour shows no correlation with other properties
such as $L_{\rm{fir}}$.

\renewcommand{\baselinestretch}{1.0}

\renewcommand{\baselinestretch}{1.0}

\subsection{\label{qS}Is there a Cold Component?}
To try to determine whether or not a cold component is necessary, we
also tried to fit isothermal SEDs to the expanded data sets for
various assumptions about $\beta$. Table~\ref{singtchiT} shows how
many good fits were produced for freely varying $\beta$ as well as for
fixed values of 1.5 and 2.0. It can be seen that if no constraints are
placed on $\beta$ then only 7 objects are incompatible with a single
temperature SED and {\em require\/} a cold component (although many
more will have better fits to a two-component model than to a single
temperature one, it is not possible to rule the single temperature
model out). The balance shifts as $\beta$ is forced to be steeper,
with only two objects allowing single temperatures for $\beta=2$. This
suggests that if we knew the value of $\beta$ (if indeed one value
should be applied to all of the sources), we could determine whether
or not a cold dust component is necessary for these galaxies.

\renewcommand{\baselinestretch}{1.0}
\begin{table}
\centering
\small
\caption{\label{singtchiT} Galaxies which could be fitted by single temperature SEDs for various $\beta$.}
\begin{tabular}{ccc}
\\[-2ex] 
\hline
\\[-2.5ex]
\multicolumn{1}{c}{$\beta$}&\multicolumn{1}{c}{Fit
($\chi_{\nu}^2<1$)}&\multicolumn{1}{c}{No fit}\\
\\[-2.5ex]
\hline
free & 25 & 7 \\
1.5 & 18 & 14 \\
2.0 & 2 & 30 \\
\hline
\end{tabular}
\end{table}

\subsection{\label{betaS}Constraints on $\beta$}
The value of $\beta$ is of great importance, as it tells us about the
behaviour of the dust emissivity with wavelength. The dust mass
opacity coefficient $\kappa_{d}$ has only been measured at FIR
wavelengths ($\sim 125\mu$m) and must be extrapolated to the
wavelength of observation as $\propto \lambda^{-\beta}$. Knowledge of
$\beta$ therefore improves our estimate of $\kappa_{d}(850)$, and
hence dust mass. A well determined $\beta$ is also necessary for
interpretation of the high redshift submm observations, where only 1
or 2 fluxes are available, making a simultaneous determination of $T$
and $\beta$ impossible and attempts to use the submm flux ratios as
redshift indicators very uncertain. However, one must also be wary
that there may be evolution in the dust properties, such as
temperature, emissivity and extinction behaviour with redshift and
metallicity. These issues are not yet fully understood, but it is
worth noting that observations of high-redshift radio galaxies and
quasars are adequately represented by an emissivity proportional to
$\nu^{1.5-2.0}$ (Benford et al. 1999; Ivison et al. 1998; Downes et
al. 1992; Hughes et al. 1997; Chini \& Kr\"{u}gel
1994). Section~\ref{2sedS} has shown that two-component fits to small
data sets with a free $\beta$ cannot tell us anything conclusive about
either the value of $\beta$ or the putative cold component. We will
now explore other ways to deduce the value of $\beta$.
\subsubsection*{Theory and observations}
A mixture of laboratory experiments and theoretical models have
produced various estimates of what $\beta$ should be for different
chemical compositions believed to represent interstellar dust,
at various temperatures and sizes. These studies are not in great
consensus. Draine \& Lee (1984) measured optical constants for a
mixture of silicates and graphite and in their model $\beta\sim2$ is
applicable for $40\leq\lambda\leq 1000\mu$m (Gordon 1995). Agladze et
al. (1996) found in laboratory experiments that there may be a
dependence of $\beta$ on $T_{\rm{d}}$ for certain types of amorphous
silicates; $\beta=2$ at 20 K and decreases with temperature although
it is still $>1.5$ at 30 K (however these measurements were made in
the millimetre regime not the submillimetre). Other samples in the
same experiment showed no such behaviour with temperature, and had a
constant value of $\beta=1.2$. Mennella et al. (1998) also found a
dependence of $\beta$ on $T$ between 20$\mu$m -- 2 mm but at higher
temperatures (24 -- 295 K). They used samples of silicates and
amorphous carbons believed to be analogous to interstellar dust. The
values of $\beta$ appropriate for the silicates in the temperature
range appropriate for ISM dust grains (10--50 K) are still clustered
around 2 without a very strong temperature trend. For the amorphous
carbon grains a low $\beta\sim1$ was favoured, while graphitic grains
had $\beta=2$ (Mennella 1995), however the low $\beta$ for the
amorphous carbons could be due to the very small particle sizes used
$\leq0.01\mu$m. Fluffy composite or fractal grains have also been
considered, which could be representative of grains found in dark
clouds. Mathis \& Whiffen (1989) found $\beta=1.5$ for composite
grains and Wright (1987) gives $0.6\leq\beta\leq1.4$ for fractal
grains. However, similar fluffy composites were investigated by
Kr\"{u}gel \& Siebenmorgen (1994) who found $\beta=2$, except for the
largest grains ($>30\mu$m) which are not believed to be representative
of general diffuse ISM dust anyway. From these studies, it is
difficult to justify a universal value of $\beta$ for different
regions in the ISM, since composition, size and possibly temperature
could have an effect.

Observations of diverse environments within the Galaxy also suggest a
range of values for $\beta$. COBE/FIRAS studies of the diffuse ISM in
the Milky Way (Sodroski et al. 1997; Reach et al. 1995; Masi et
al. 1995) find $\beta=1.5-2.0$ depending on whether one or two
temperature components are fitted (a two-component model with
$\beta=2$ gives the best fit). Warm clouds associated with H{\sc{ii}}
regions are also found to have $\beta=2$ (Gordon 1988). Recent results
from the submm balloon experiment {\em PRONAOS\/} (Dupac et al. 2001)
find tentative evidence for an inverse dependence of $\beta$ on $T$
but the result is largely based on one cloud (OMC-1) having a lower
$\beta$ and higher temperature than the others. There are problems
with this interpretation if OMC-1 contains dust at more than one
temperature (i.e. the same problem we have here), also the
temperatures of the clouds investigated by Gordon (1988) ranged from
30--80 K, yet all apart from OMC-1 had $\beta=2$. It may be
that OMC-1 contains a grain population with larger than normal sizes, or more
than usual amounts of a particular grain composition. Colder clouds in
Orion A have also been observed, and $\beta$ found to be 1.9
(Ristorcelli et al. 1998). However, observations of both young and
evolved stars have revealed low $\beta\sim0.2-1.4$ (Weintraub et
al. 1989; Knapp et al. 1993) and this is usually attributed to grains
growing to very large sizes in cool, dusty envelopes or disks.

In contrast, multi-wavelength observations of external galaxies are
almost universal in their finding that $1.5\leq\beta\leq2$ and
favouring $\beta=2$ (Chini et al. 1989; Chini
\& Kr\"{u}gel 1993; Alton et al. 1998; Frayer et al. 1999; Bianchi et
al. 1998; Braine et al. 1997). It may be that the contribution to the
global SED from dust around stars (which displays low $\beta$) is
small when averaged over the whole galaxy. Assuming that the
production of dust occurs in similar ways in all galaxies (analogous
to assuming star formation mechanisms are universal) then maybe we
{\em should\/} expect the global dust properties to be generally
consistent from galaxy to galaxy (after all we must make some such
assumption if we wish to compare dust masses etc).

\subsubsection*{Colour plots: the ratio $S_{450}/S_{850}$}

\begin{figure*}
\centerline{\psfig{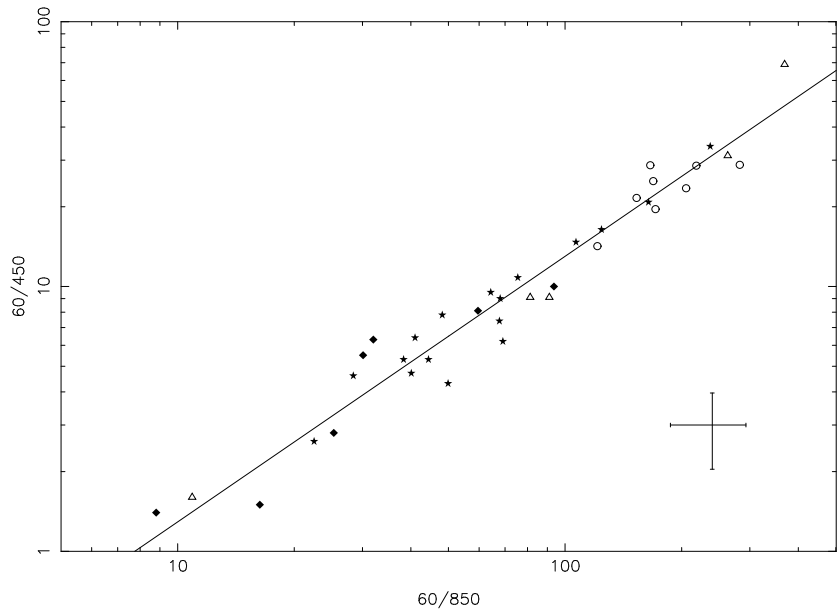}}
\caption[Colour-colour plot for the objects in the
IRAS sample with 450$\mu$m fluxes.]{\label{450850F} Colour-colour plot
for the objects in the IRAS sample with 450$\mu$m fluxes (stars and
circles), other objects from the literature including Mrk 231
(triangles) and some of the optically selected sample (diamonds). The
conservative uncertainty in the ratios is shown based on flux error of
10\% for 60$\mu$m, 20\% for 850$\mu$m and 30\% for 450$\mu$m. The best
fit line is shown, which has a slope of unity. Data for the stars and
open circles is from this paper, triangles from Rigopoulou et
al. (1996); Alton et al. (1998); Frayer et al. (1999), and the
diamonds are from Dunne (2000)}
\end{figure*}

Figure~\ref{450850F} shows the colour-colour plot for $S_{60}/S_{450}$
versus $S_{60}/S_{850}$ for the objects in this sample which have
450$\mu$m fluxes (stars and circles) plus some objects from the
literature (which are not in our sample) with fluxes at these
wavelengths (IC 1623, NGC 891, Mrk 231, Mrk 273, UGC 5101 -- open
triangles) plus some of the optically selected galaxies from our as
yet incomplete sample (diamonds). The stars refer to objects which we
observed and reduced, and the circles are 450$\mu$m fluxes from the
literature for this sample.

\renewcommand{\baselinestretch}{1.0}

There are a few points to note here:
\begin{enumerate}
\item{The relationship is very tight (correlation coefficient
$r_s=0.96$, significance=$5.8\sigma$) and the scatter is consistent
with the uncertainties in the flux ratios (denoted on the
figure).} 
\item{This relationship holds across 2--3 orders of magnitude in
$L_{\rm{fir}}$, even though there are many different types of
galaxies, ranging from optically selected galaxies with low $L_{\rm{fir}}$
to very active ULIRGS and AGN such as Arp 220 and Mrk 231.}
\item{The best fitting line (least squares) is:
\begin{equation}
\log(S_{60}/S_{450})=(1.01\pm0.03)\,\log(S_{60}/S_{850})-(0.909\pm0.045)
\end{equation}
This can be re-written as $S_{60}/S_{450}
=0.123\,(S_{60}/S_{850})^{1.01}$ which implies that, within the
uncertainties, the ratio $S_{450}/S_{850}$ is constant. From the
fluxes, the mean $S_{450}/S_{850}=7.9$ with $\sigma=1.6$ which is a
remarkably tight distribution, consistent with the errors on the fluxes.}
\end{enumerate}

\begin{figure*}
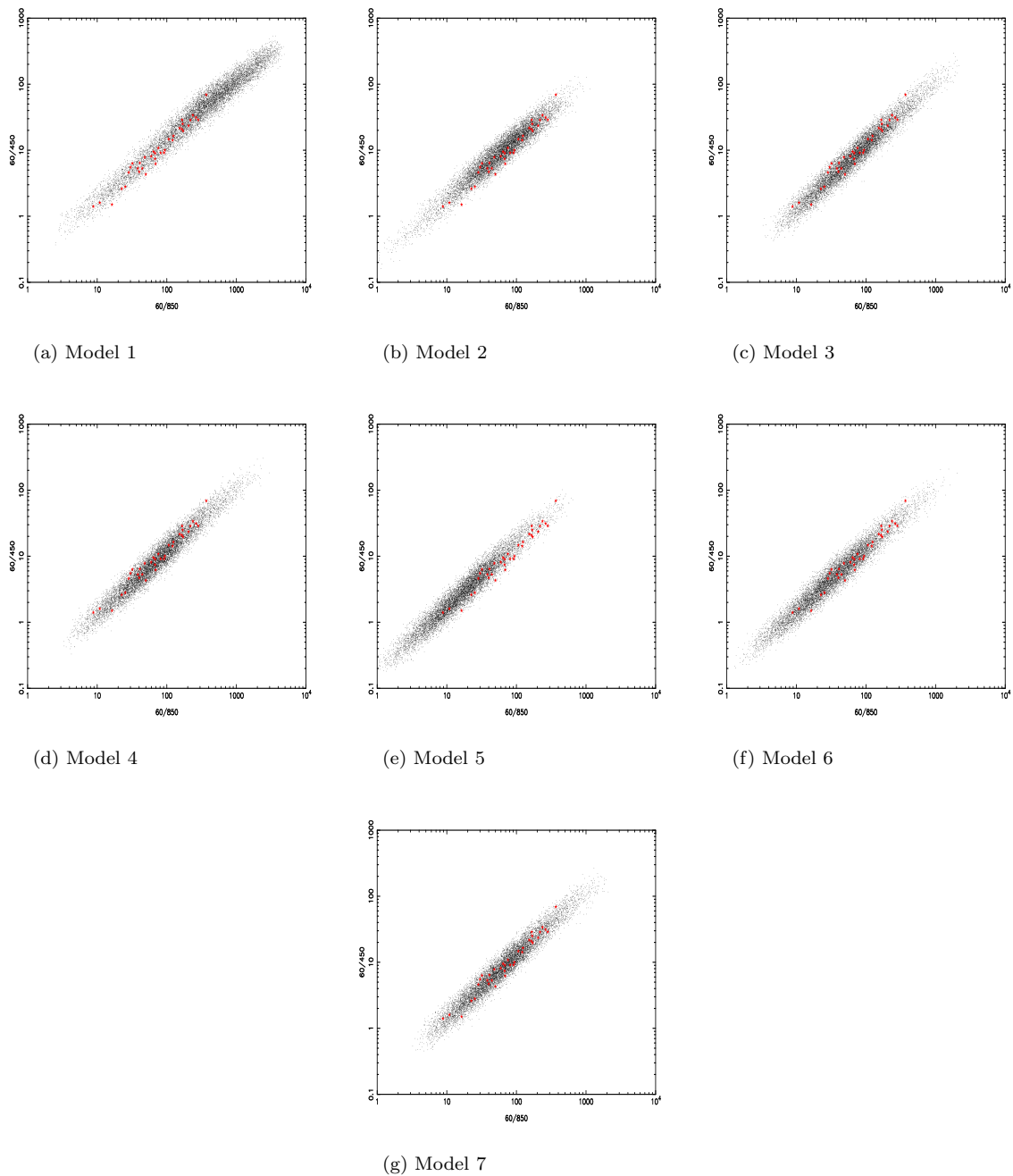

 \subfigure[Model 1]{\psfig{file=Fig7a.ps,height=4.5cm,width=4.5cm,angle=-90}}\goodgap
 \subfigure[Model 2]{\psfig{file=Fig7b.ps,height=4.5cm,width=4.5cm,angle=-90}}\goodgap
 \subfigure[Model
3]{\psfig{file=Fig7c.ps,height=4.5cm,width=4.5cm,angle=-90}}\\
 \subfigure[Model 4]{\psfig{file=Fig7d.ps,height=4.5cm,width=4.5cm,angle=-90}}\goodgap
\subfigure[Model 5]{\psfig{file=Fig7e.ps,height=4.5cm,width=4.5cm,angle=-90}}\goodgap
 \subfigure[Model
6]{\psfig{file=Fig7f.ps,width=4.5cm,height=4.5cm,angle=-90}}\\
 \subfigure[Model
7]{\psfig{file=Fig7g.ps,width=4.5cm,height=4.5cm,angle=-90}}\\
\caption{\label{pointsF} Comparison of the 10,000 simulated flux
ratios and the actual data points for the 37 galaxies (red stars) for
models 1--7.}
\end{figure*}

What does a constant value of $S_{450}/S_{850}$ mean? The flux ratio
can be expressed as a function of $T$ and $\beta$ as follows:
\begin{equation}
\frac{S_{450}}{S_{850}}=\left(\frac{\nu_{450}}{\nu_{850}}\right)^{\beta}\times\frac{{\rm{B}}(\nu_{450}, T)}{{\rm{B}}(\nu_{850}, T)}
\label{4508502E}
\end{equation}
Since the submm fluxes are on the Rayleigh-Jeans side of the Planck
function the temperature dependence is not very steep, roughly
$\propto T$. A constant value for the ratio therefore implies a small
range of real $\beta$ and $T_c$ in the ISM of galaxies. We further
investigated the implications of the relationship in
Fig.~\ref{450850F} in the following way: Hypothetical SEDs were
created using the parameters $T_w$, $T_c$, $N_c/N_w$ and $\beta$, the
values for these being either specified, drawn from a uniform random
distribution between two given limits, or drawn from a Gaussian
distribution with a designated mean and dispersion. The various models
are listed in Table~\ref{SEDmodsT}. Each SED was used to produce
fluxes at 60, 450 and 850$\mu$m. These fluxes were then subjected to
`observational uncertainties' by drawing a new flux from the Gaussian
distribution with the original flux as the mean and 1$\sigma$
uncertainties derived from the observations as 7, 25 and 14 percent at
60, 450 and 850$\mu$m respectively. Next, only those new fluxes which
were found to be $\geq 3$ times the error were selected, to mimic
detection limits. The required flux ratios $S_{60}/S_{450}$,
$S_{60}/S_{850}$ and $S_{450}/S_{850}$ were then calculated from these
final fluxes. In order to be able to compare such things as the linear
fitting parameters to those derived from the actual data, we needed to
use the same number of data points (37). Therefore fluxes were created
in sets of 37, and each set was fitted in exactly the same way as the
real data (simple least squares analysis) and the standard deviation
on the ratio $S_{450}/S_{850}$ calculated. 10,000 versions of
Fig.~\ref{450850F} were created for each SED model, and used to
produce an estimate of the mean slope and intercept of the best
fitting line and also $\sigma_{450/850}$ (the dispersion on the ratio
$S_{450}/S_{850}$). Additionally, one point from each set of 37 was
taken and placed into a separate group. These 10,000 flux ratios were
then compared with the distribution of the real flux ratios via
Kolmogorov-Smirnoff two-sample tests. These points are plotted against
the real data for the different models in Fig.~\ref{pointsF}.  We
tried different models to see what combination of parameters best
reproduced the statistical properties of the data, in particular:
\begin{enumerate}
\item{the mean values and distributions of the three flux ratios.}
\item{the best fitting slope and intercept.}
\item{the dispersion of the flux ratio $S_{450}/S_{850}$.}
\end{enumerate}
The models represent a range of `realistic' isothermal and
two-component SEDs of the form fitted to the fluxes earlier. One
additional model was included to account for the possible contribution
of the $^{12}$CO(3--2) line to the 850$\mu$m filter. This line is excited in
warm ($T_k>33$ K), dense gas and may comprise a non-negligible
fraction of the bolometer flux. Observations of this line are still
relatively rare as they are hampered by the same difficult observing
conditions which have held continuum observers in the dark for so
long. The only accurate way to ascertain the contribution of the line
to the 850$\mu$m flux is to map the same galaxies in CO(3--2) and
compare the total flux in the line to that in the continuum. Failing
this, rough estimates of the CO(3--2) contribution may be made in a
variety of ways (see Dunne 2000 for details, also Papadopoulos \&
Allen 2000; Israel et al. 1999; Bianchi et al. 1998; Mauersberger et
al. 1999), all consistent with approximately 4--30 percent of the
total continuum flux being due to CO(3--2) line emission. This large
range reflects the strong dependence of CO(3--2) on the physical
conditions in the gas. CO(3--2) observations of 50 galaxies from the
SLUGS IRAS sample are in progress which should allow us to make a
better estimate of the contamination in the future (Seaquist, private
communication). The effects of possible line contamination are not
very large (as the suspected line contribution is comparable to the
uncertainty in the submm fluxes) but would lead to us overestimating
the dust masses by up to 30 percent and underestimating $\beta$ by
$\sim 0.1$ in both the isothermal and two-component temperature cases.
To mimick a CO(3--2) contribution Model 7 simply adds 10 percent to
the predicted 850$\mu$m flux from the SED template, which has the
effect of reducing the flux ratios for a given $\beta$. The tightness
of the correlation in Fig.~\ref{450850F} suggests that the
contribution by this line adds little to the scatter and is therefore
either relatively small or a remarkably constant fraction of the
continuum flux. The results of the comparisons of simulations and data
are in Table~\ref{simresT}.

\renewcommand{\baselinestretch}{1.0}
\begin{table*}
\centering
\caption{\label{SEDmodsT} Parameters for SED templates. }
\begin{tabular}{ccccc}
\\[-2ex] 
\hline
\\[-2.5ex]
\multicolumn{1}{c}{Model}&\multicolumn{1}{c}{$T_w$
(K)}&\multicolumn{1}{c}{$T_c$ (K)}&\multicolumn{1}{c}{$\beta$}&\multicolumn{1}{c}{$N_c/N_w$}\\
\\[-2.5ex]
\hline
1 & Uni 25 -- 65 & -- & Uni 1--2 & -- \\
2 & Gau $\mu =35.7$, $\sigma =5.3$ & -- & Gau $\mu=1.3$,
$\sigma=0.2$ & --\\
3 & Uni 30--55 & Gau $\mu=20$, $\sigma=2.5$ & 2 & Uni
1--100\\
4 & Uni 30--55 & 18 & 2 & Uni 1--100\\
5 & Uni 30--55 & Uni 15--25 & 1.5 & Uni 1--100\\
6 & Uni 30--55 & Gau $\mu=20$, $\sigma=2.5$ & Uni 1.5--2.0 
& Uni 1--100\\
7$^{*CO}$ & Uni 30--55 & Gau $\mu=20$, $\sigma=2.5$ & 2 & Uni 1--100\\
\hline
\end{tabular}
\flushleft
\small (1) Model number. (2) Warm temperature distribution. Uni = 
a uniform random selection between the two limits, Gau = drawn from a
Gaussian distribution with mean $\mu$ and dispersion $\sigma$. (3)
Cold temperature distribution. No entry indicates an isothermal
model. (4) Emissivity index. (5) Distribution of ratio of cold-to-warm
dust.\\ Model 2 uses the mean values and dispersions in $T_{\rm{d}}$
and $\beta$ from the isothermal fits in Paper I.\\Model 7$^{*CO}$ has
the SED parameters listed but then adds a further 10 percent to the
predicted 850$\mu$m flux to mimic a contribution from the CO(3--2)
line.
\end{table*}

\renewcommand{\baselinestretch}{1.0}
\begin{table*}
\centering
\caption{\label{simresT} Results of simulations.}
\begin{tabular}{cccccccccc}
\\[-2ex] 
\hline
\\[-2.5ex]
\multicolumn{1}{c}{(1)}&\multicolumn{1}{c}{(2)}&\multicolumn{1}{c}{(3)}&\multicolumn{1}{c}{(4)}&\multicolumn{1}{c}{(5)}&\multicolumn{1}{c}{(6)}&\multicolumn{1}{c}{(7)}&\multicolumn{1}{c}{(8)}&\multicolumn{1}{c}{(9)}&\multicolumn{1}{c}{(10)}\\
\multicolumn{1}{c}{Model}&\multicolumn{1}{c}{Slope}&\multicolumn{1}{c}{Int.}&\multicolumn{1}{c}{$\sigma_{450/850}$}&\multicolumn{1}{c}{$\langle
S_{60}/S_{450}\rangle$}&\multicolumn{1}{c}{KS}&\multicolumn{1}{c}{$\langle
S_{60}/S_{850}\rangle$}&\multicolumn{1}{c}{KS}&\multicolumn{1}{c}{$\langle
S_{450}/S_{850}\rangle$}&\multicolumn{1}{c}{KS}\\
\hline
\\[-2.5ex]
Data & $1.01\pm 0.03$ & $-0.909\pm 0.045$ & $1.6^{+0.42}_{-0.24}$ & $13.9\pm 2.1$ & & $104
\pm 14$ & & $7.90 \pm 0.26$ &\\
\\[-2.0ex]
1 & 0.915 & $-0.684$ & $2.67$ & 64.6 &
0.52 & 574 & 0.49 & 8.13 & 0.15 \\
  & $3.2\sigma$ & $5\sigma$ & 2.5e-4 &  & 2.4e-9 &
 & 1.9e-8 &  & 0.33\\
\\[-2.0ex]
2 & 0.924 & $-0.680$ & $2.01$ & 12.9 & 0.1 & 89.9 & 0.13 & 6.80 & 0.32\\
 & $2.9\sigma$ & $5.1\sigma$ & 0.079 &  & 0.83 &
 & 0.51 &  & 5e-4 \\
\\[-2.0ex]
3 & 0.993 & $-0.910$ & $2.36$ & 14.6 & 0.16 & 124.4 & 0.09 & 8.72 & 0.23 \\
 &  $0.6\sigma$ & $0.02\sigma$ & 4.1e-3 &  & 0.26 &
 & 0.89 &  & 0.03\\
\\[-2.0ex]
4 & 0.987 & $-0.881$ & $2.22$ & 15.9 & 0.18 & 133.2 & 0.14 & 8.34 & 0.17 \\
 & $0.8\sigma$ & $0.6\sigma$ & 0.0137 &  & $0.18$ &
 & 0.46 &  & 0.23\\
\\[-2.0ex]
5 & 0.992 & $-0.774$ & $1.73$ & 5.5 & 0.5 & 34.2 & 0.55 & 6.32 & 0.42\\
 & $0.6\sigma$ & $3.0\sigma$ & 0.521 &  & 8.7e-9 &
 & 2e-10&  & 2e-6\\
\\[-2.0ex]
6 & 0.970 & $-0.808$ & $2.14$ & 9.0 & 0.33 & 68.3 & 0.33 & 7.47 & 0.20\\
 & $1.3\sigma$ & $2.2\sigma$ & 0.026 &  & 4.3e-4 &
 & 5.7e-4 &  & 0.089\\
7$^{*CO}$ & 0.992 & $-0.869$ & $2.15$ &
14.5 & 0.16 & 113.3 & 0.13 & 7.92 & 0.12\\
 & $0.6\sigma$ & $0.9\sigma$ & 0.022 &  & 0.26 &
 & 0.56 &  & 0.69\\
\hline
\end{tabular}
\flushleft
\small (1) Model name, first row is the actual data. Uncertainties in
the values are from the least-squares linear fitting and from the
observed dispersions in the flux ratios. The quoted error on
$\sigma_{450/850}$ is a 95\% confidence interval. The errors on the
flux ratios are the errors on the mean value. (2) Gradient of the
relationship $\log(S_{60/450}), \log(S_{60/850})$ as in
Fig.~\ref{450850F}. Determined by simple linear regression (least
squares) on sets of 37 points. (3) Intercept of the same
relationship. For the models, the value is the mean for
the 10,000 simulations. (4) Dispersion in the ratio of
$S_{450}/S_{850}$. (5,7,9) Overall mean ratio of fluxes (10,000
points). (6,8,10) Result of a two sample KS test on the 37 data points
and the 10,000 simulated ratios.\\ The significance of the differences
between models and data are listed beneath each entry, either as
no. of S.D. between means for the fitted slopes and intercepts, as the
probability that the two samples have the same variance for
$\sigma_{450/850}$ (calculated from the F-distribution with 36 and
9999 degrees of freedom for the two samples), or for the KS tests, as
the probability that the two distributions are from the same parent
population.
\end{table*}

\renewcommand{\baselinestretch}{1.0}

The two isothermal SED models (1 \& 2) are not compatible with the
data as firstly, the ratio of $S_{450}/S_{850}$ is not predicted to
be constant (slope not unity), and secondly the means and distributions
of the flux ratios are not compatible with the data and are ruled out
by the KS tests at a high level of significance. Not all two-component 
models can reproduce the observed Fig.~\ref{450850F} either. Models 5
\& 6 which have lower $\beta$ values (either $\beta=1.5$ or $\beta$ has a
range of values between 1.5--2.0) are also not compatible because
their flux ratio predictions are too low (i.e. too much submm flux
compared to that at 60$\mu$m), however they do produce the correct
slope. The three models which had $\beta=2$ (3, 4 \& 7) all produced
acceptable matches to the data (the only marginal point being the
ratio of $S_{450}/S_{850}$ in Model 3 which had a KS significance of
0.03). Lowering the cold temperature slightly as in Model 4, solved
this discrepancy. Also, Model 7, which includes the effect of a
CO(3--2) contribution can produce the correct flux ratio with the same
cold temperature as Model 3 (i.e. 20 K). While there is an element of
$T_c$, $\beta$ degeneracy in fitting the ratio $S_{450}/S_{850}$ (we
could raise the cold temperature and lower $\beta$ to produce the same
ratio), this does not solve the problem with the IRAS/submm ratios
which were all significantly different from the data in the lower
$\beta$ models. For this reason, we believe that Fig.~\ref{450850F} is
evidence in favour of a steep and universal $\beta$ value close to
2. Any CO(3--2) contribution will only strengthen this conclusion and
further reject lower $\beta$ values. The warm temperatures used in the
models were based on the values which came out of the two-component
fits, and were kept constant for all models. We did experiment with
altering them but for all realistic scenarios the warm temperature
values did not have a significant effect on the models. The same is
true for the ratio of cold/warm dust, which was allowed to vary
between 1--100 as per the two component fits. The tightness of the
ratio $S_{450}/S_{850}$ was very difficult to reproduce with the
simulations, and indeed it was not obvious which factors directly
affected this. In all cases the models could not reproduce as tight a
distribution as the actual data. This may be because we have been too
conservative with our estimates of the observational uncertainties, or
more likely, because there are intrinsic relationships between such
parameters as $T_w,\, T_c$ and $N_c/N_w$ which were not accounted for
in our models.
  
A cold component temperature between 15--25 K and $\beta\sim2$ is in
good agreement with the measurements of the diffuse ISM in the Galaxy
($T\sim 17$ K, $\beta=2.0$) by COBE/FIRAS (which had enough spectral
resolution to do it properly), and with calculations of the predicted
equilibrium temperature for grains in the Galactic ISRF which is
25--15 K, depending on Galactocentric distance (Cox, Kr\"{u}gel \&
Mezger 1986).

\subsubsection*{The Verdict}
From SED fitting alone, it is often not possible to simultaneously
determine $\beta$ {\em and\/} to show conclusively that there is a
cold dust component, even when as for Arp220 (Fig.~\ref{A220F}), there
are measurements at many FIR/submm wavelengths. The results of the
simulations of the 450/850$\mu$m ratio strongly favour $\beta=2$, and
for the 7 galaxies where a cold component is essential
(Section~\ref{qS}), this is also the $\beta$ value from the fit. We
will therefore assume that $\beta=2$ is the {\em true\/} value and
that it holds for all galaxies. Since there is abundant evidence in
the literature for both cold components and $\beta\sim 2$ we accept it
as the best physical model for the SED. Acceptable two-component fits
using $\beta=2$ were possible for all of the sources.

\section{Discussion}

\begin{figure*}
 \subfigure[ugc903: $T$(45,21), n=91]{\psfig{file=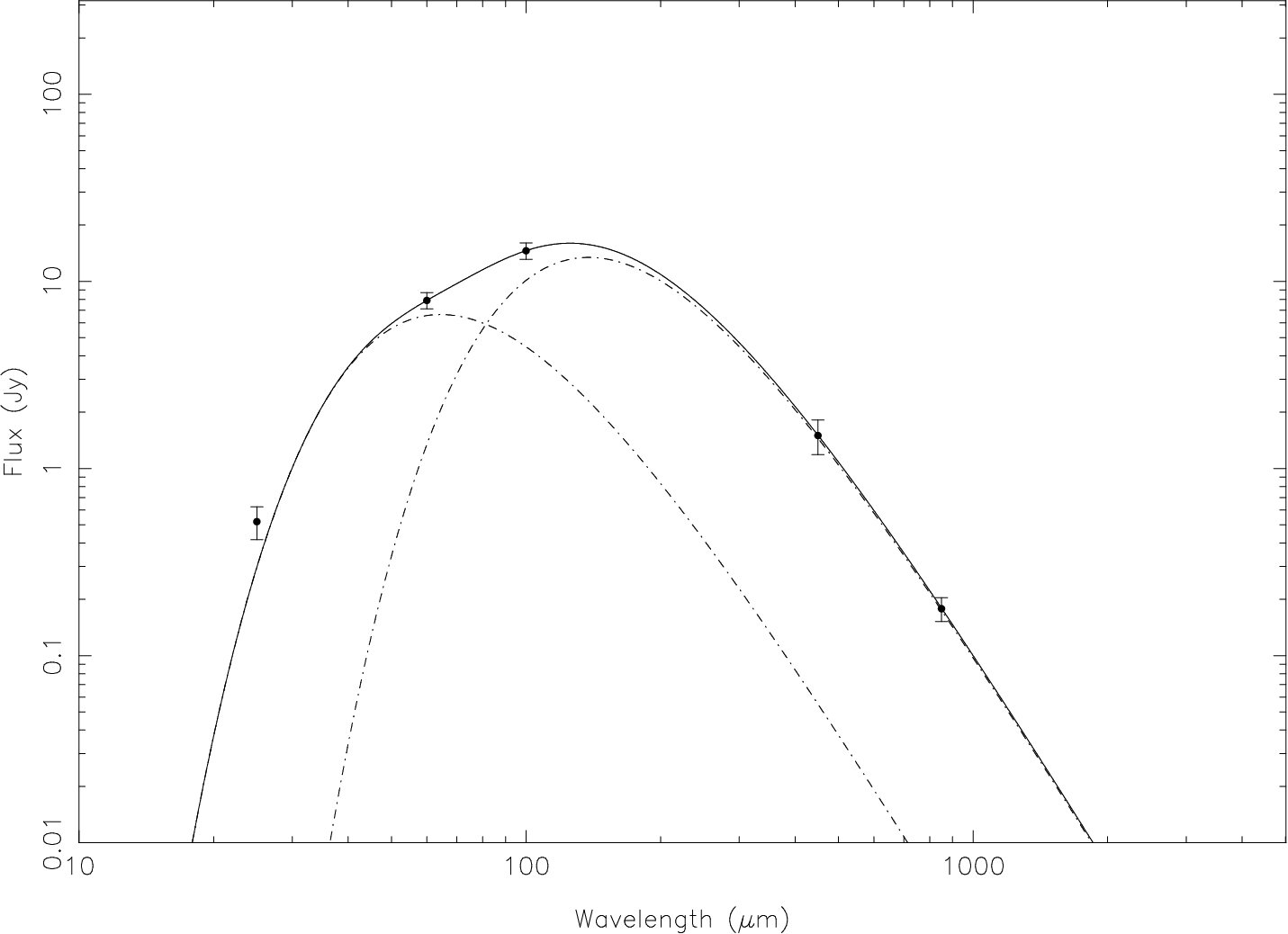,width=4.5cm,height=4cm,angle=-90}}\goodgap
 \subfigure[ngc958: $T$(44,20), n=186]{\psfig{file=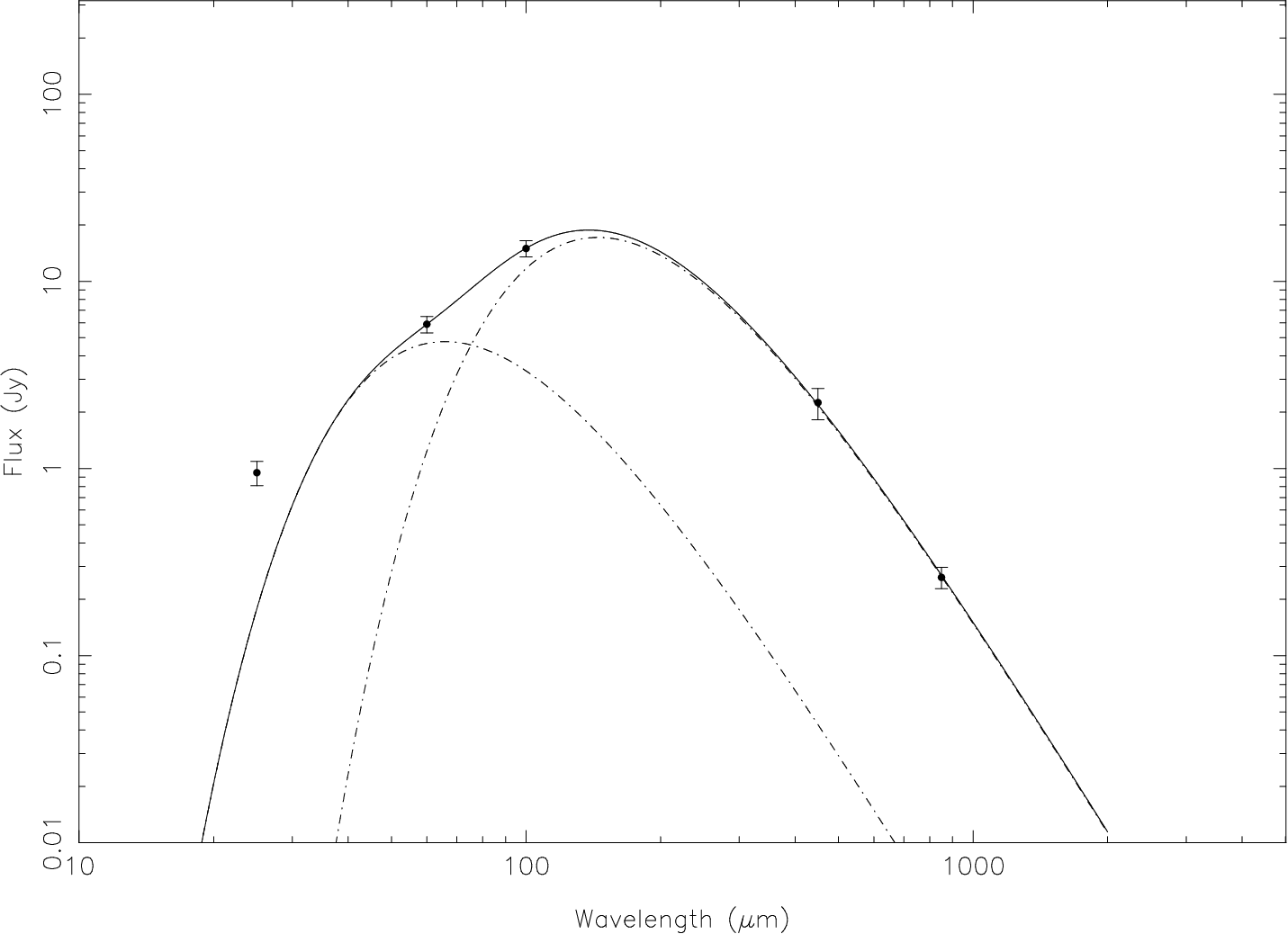,width=4.5cm,height=4cm,angle=-90}}\goodgap
 \subfigure[ugc2369: $T$(36,21), n=6]{\psfig{file=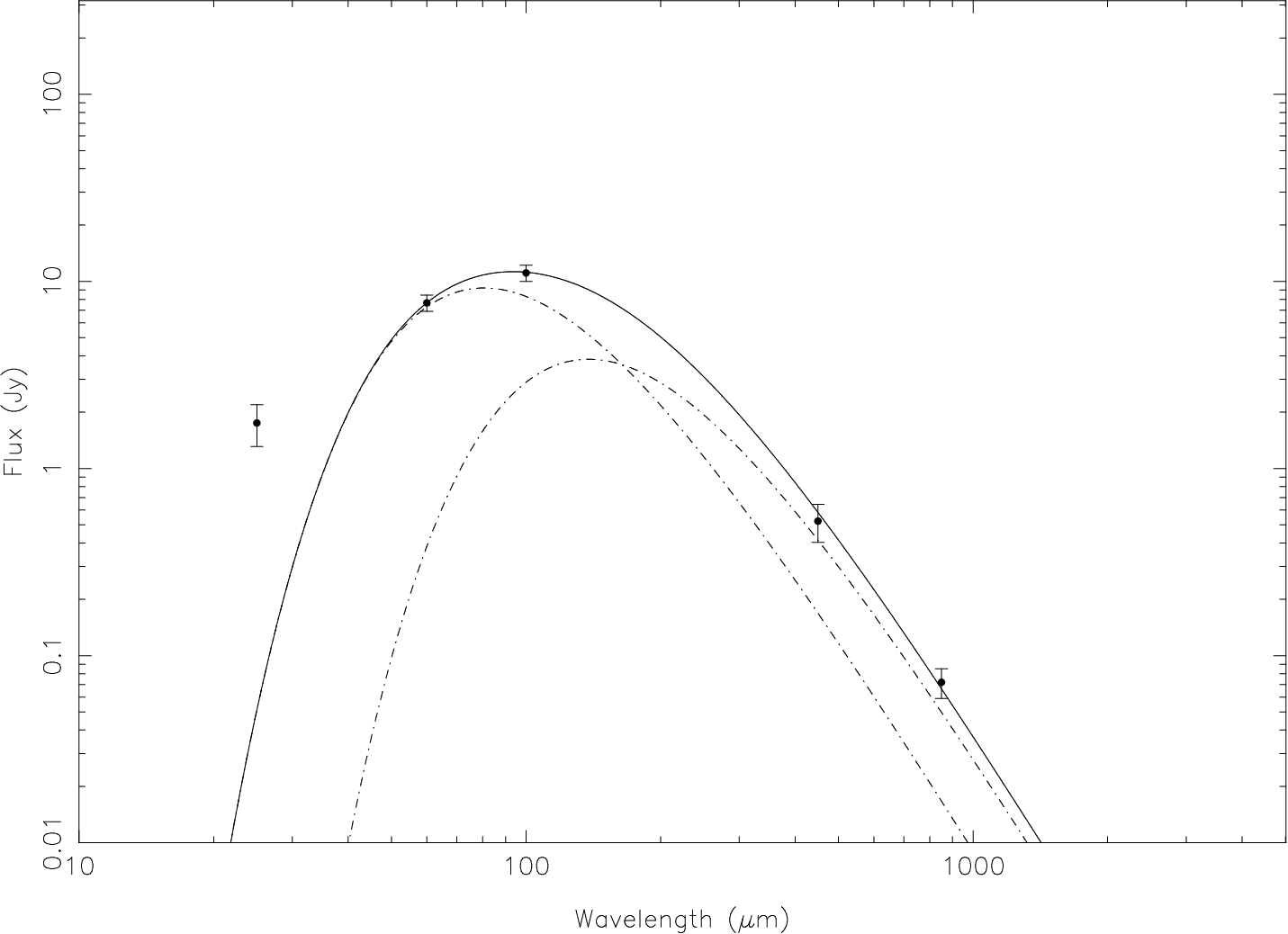,width=4.5cm,height=4cm,angle=-90}}\\
 \subfigure[ugc2403:$T$(50,22), n=100]{\psfig{file=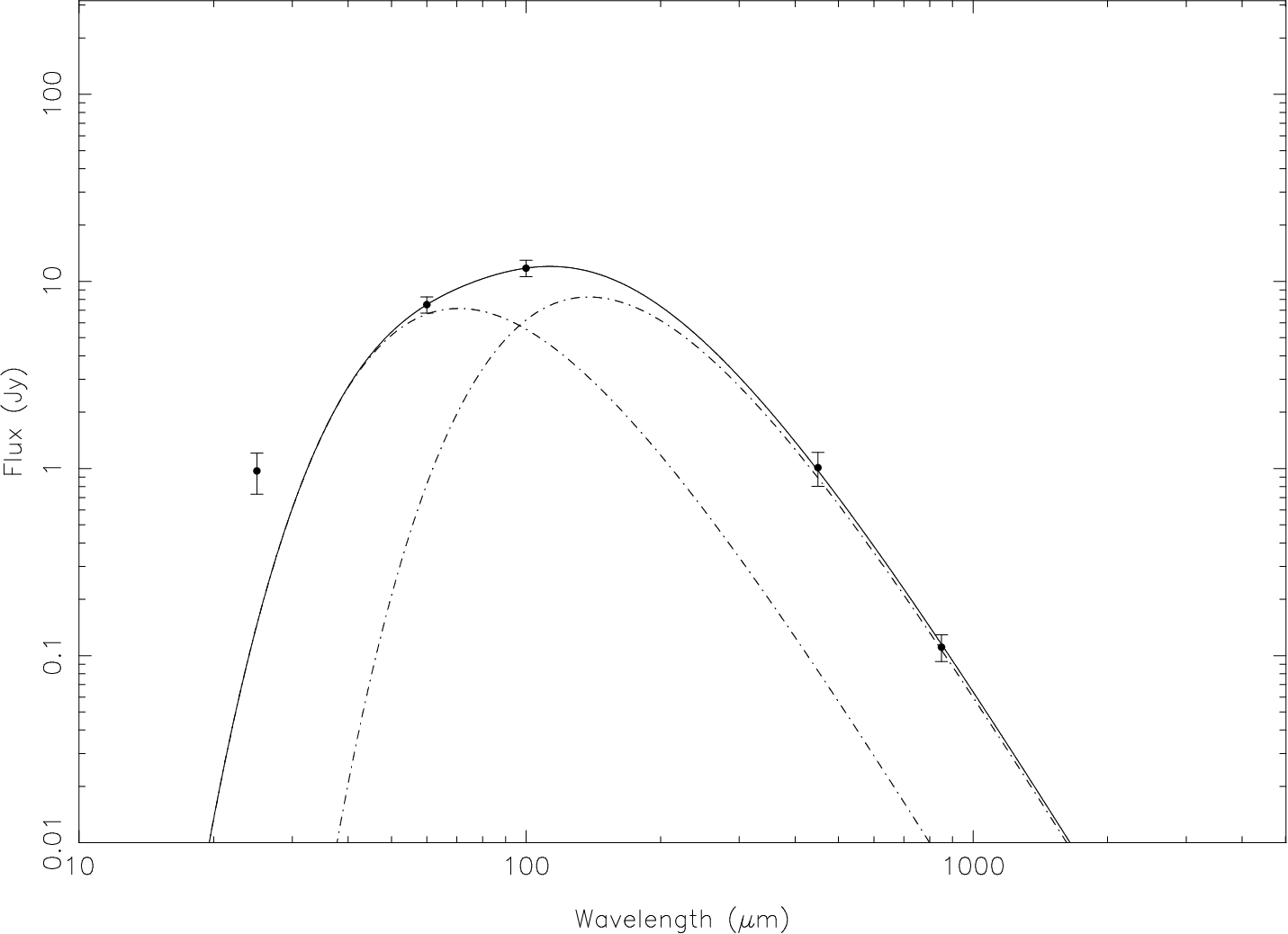,width=4.5cm,height=4cm,angle=-90}}\goodgap
 \subfigure[ir1211: $T$(49,24), n=27]{\psfig{file=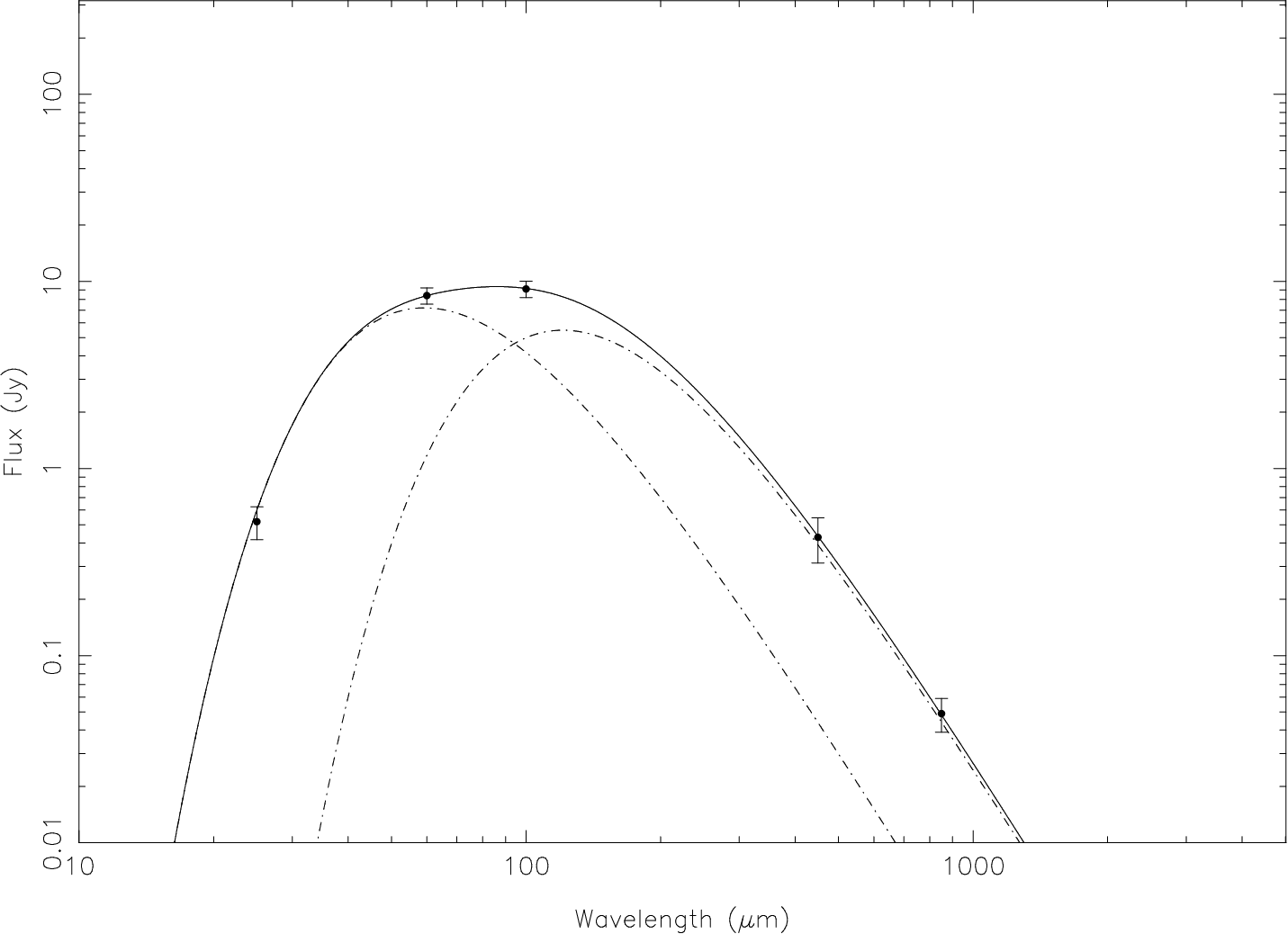,width=4.5cm,height=4cm,angle=-90}}\goodgap
 \subfigure[ngc1667: $T$(28,17), n=3]{\psfig{file=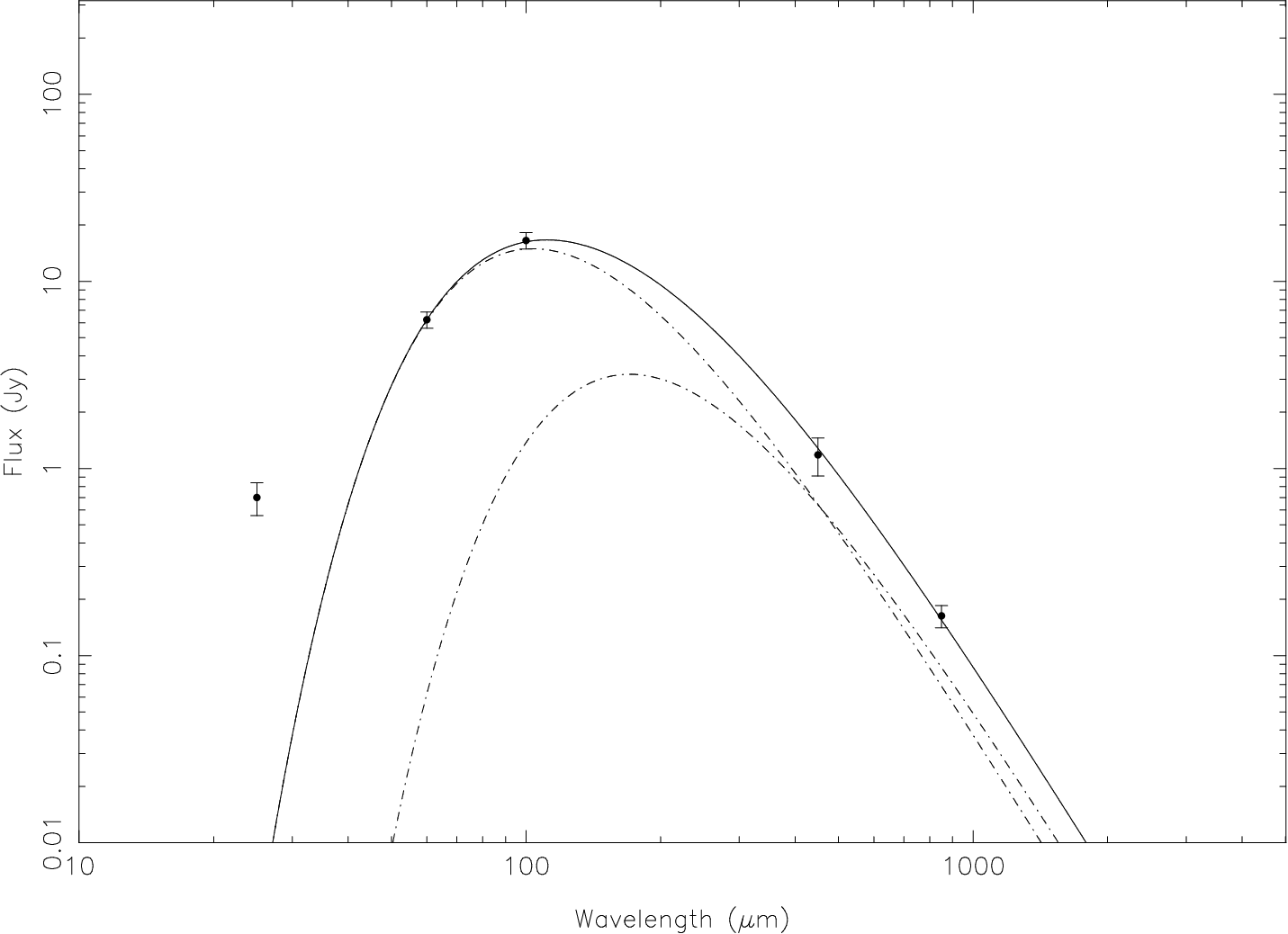,width=4.5cm,height=4cm,angle=-90}}\\
 \subfigure[ugc2982: $T$(31,17), n=6]{\psfig{file=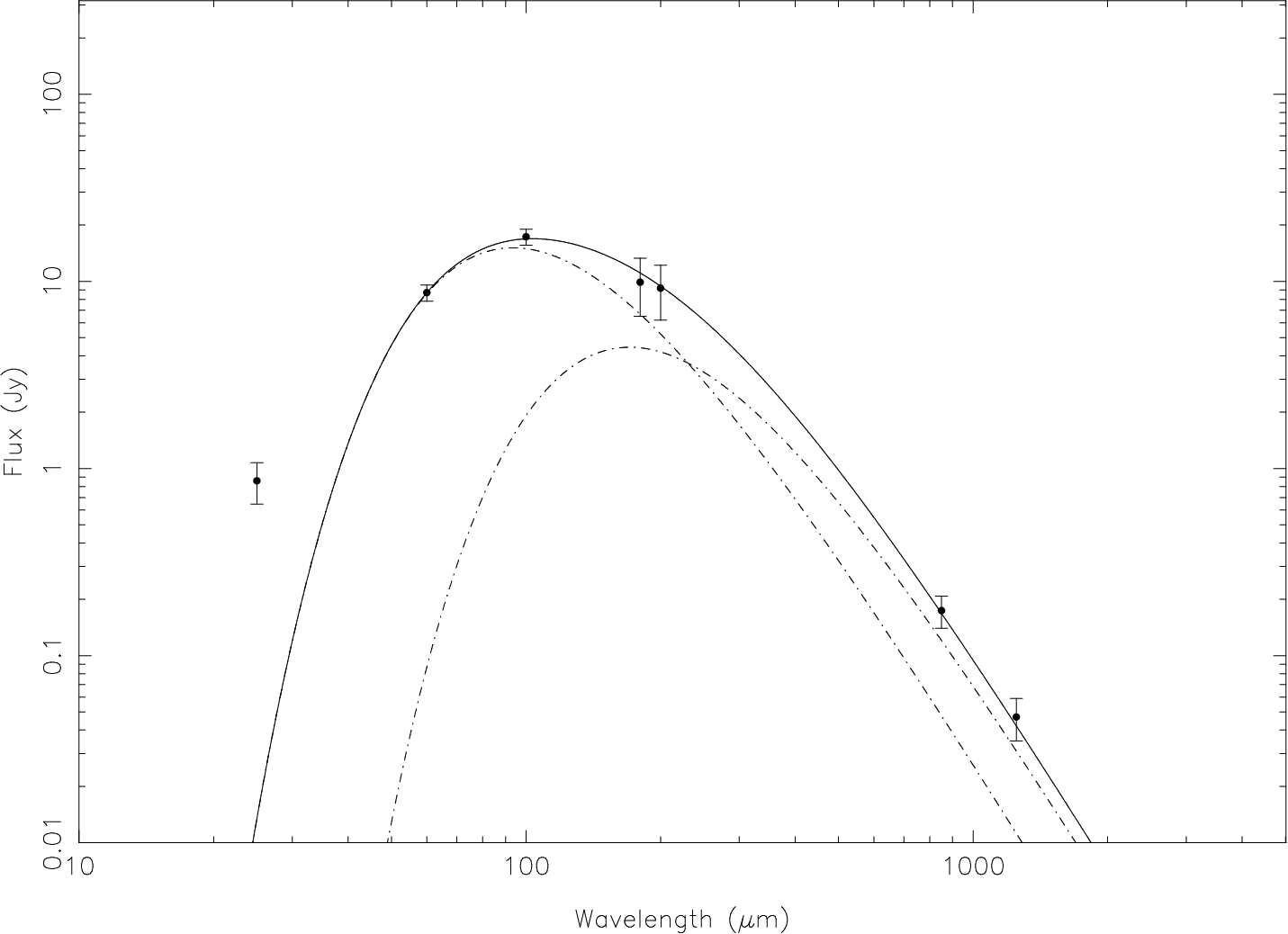,width=4.5cm,height=4cm,angle=-90}}\goodgap
 \subfigure[ir1525: $T$(45,19), n=15]{\psfig{file=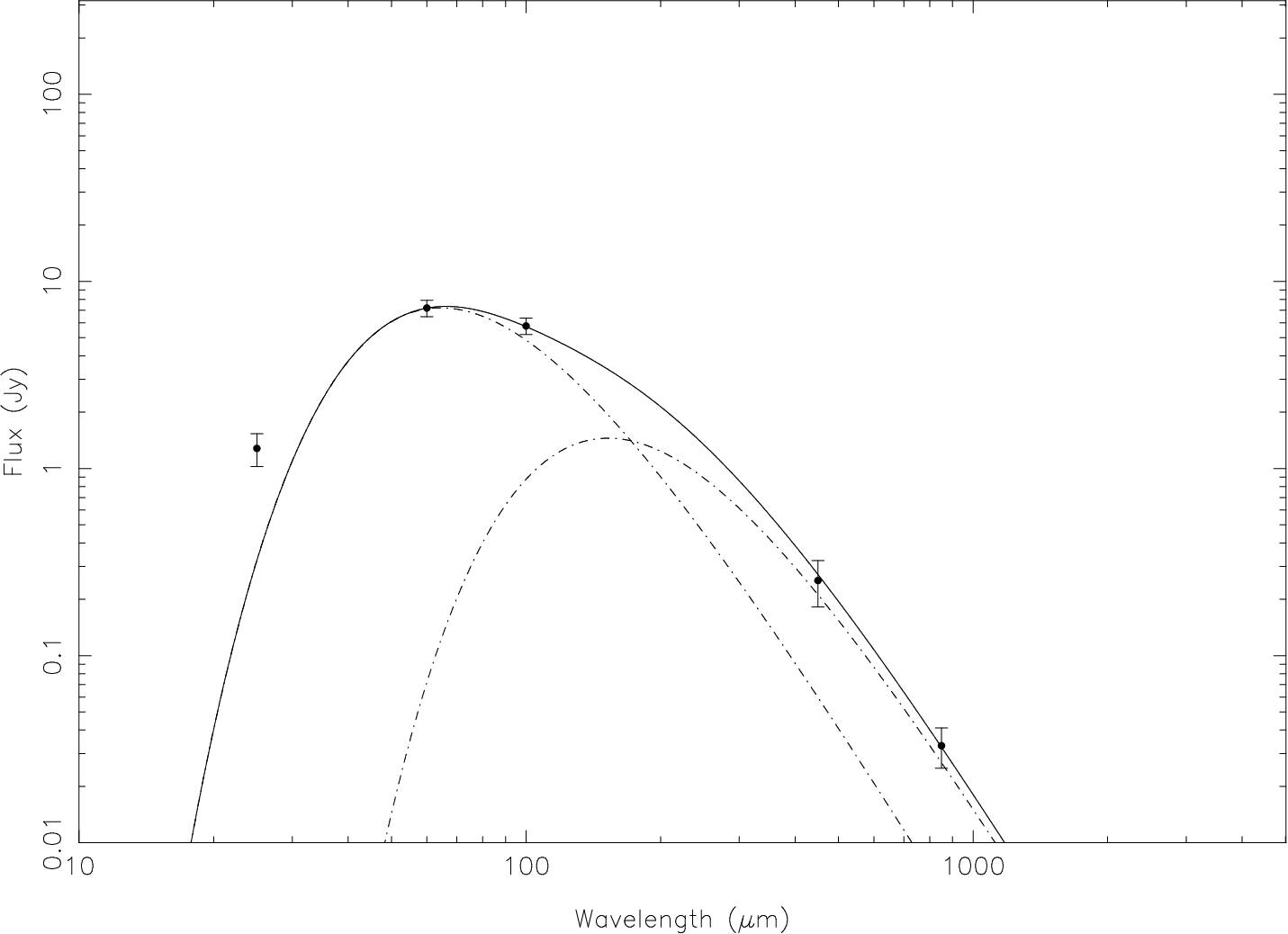,width=4.5cm,height=4cm,angle=-90}}\goodgap
 \subfigure[ugc5376: $T$(44,21), n=87]{\psfig{file=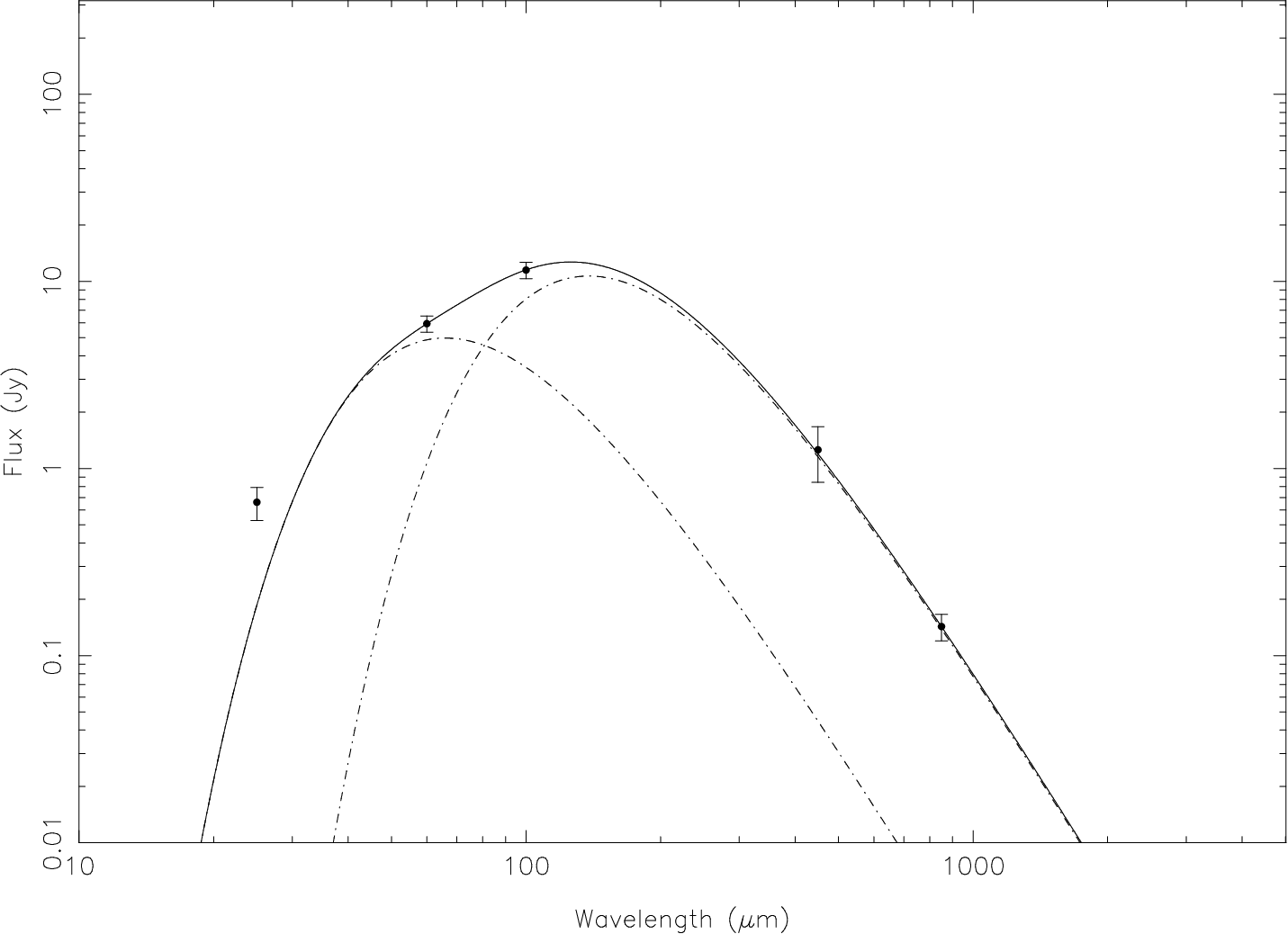,width=4.5cm,height=4cm,angle=-90}}\\
 \subfigure[ngc2623: $T$(50,27), n=18]{\psfig{file=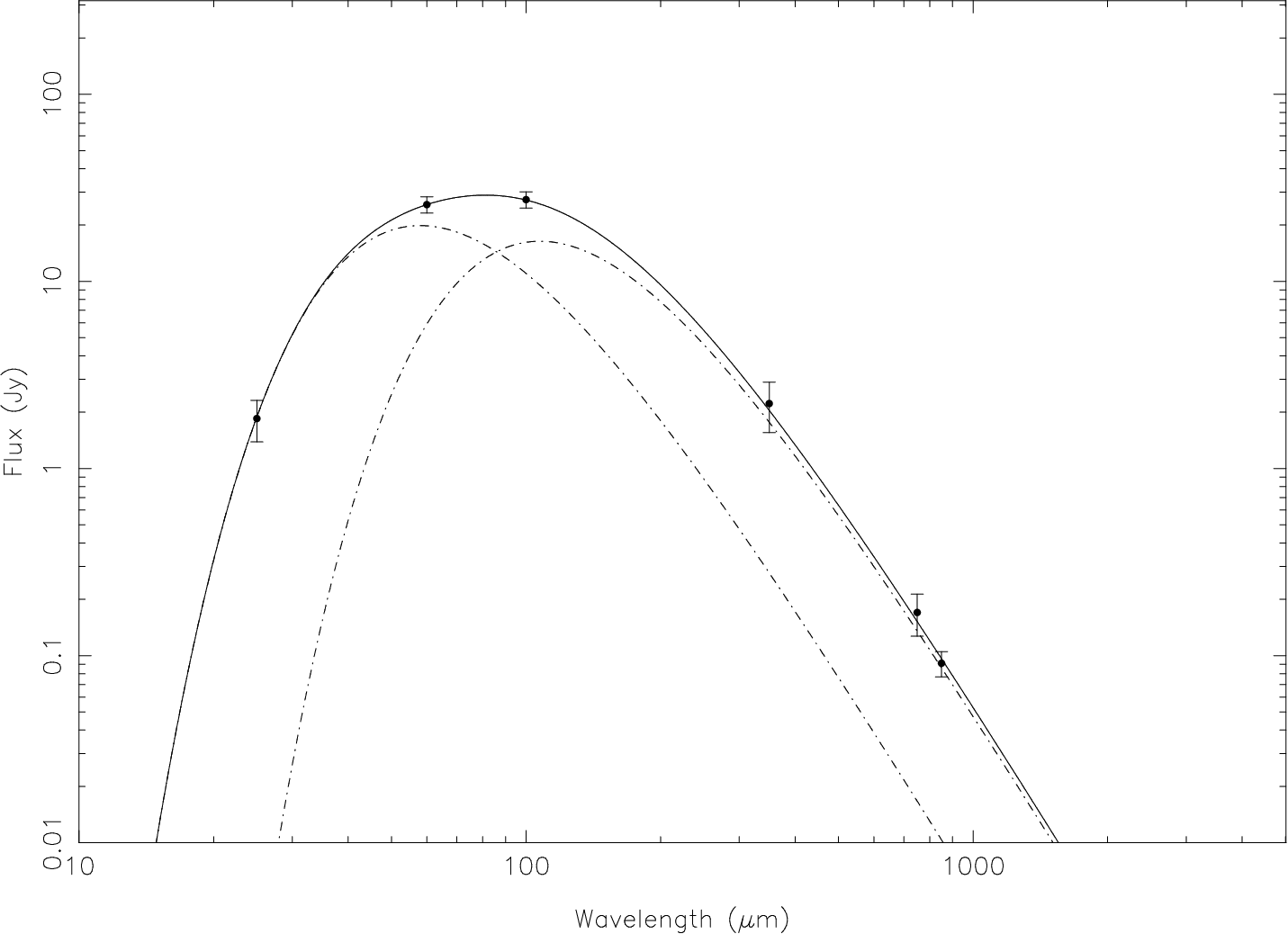,width=4.5cm,height=4cm,angle=-90}}\goodgap
 \subfigure[arp220: $T$(48,18), n=42]{\psfig{file=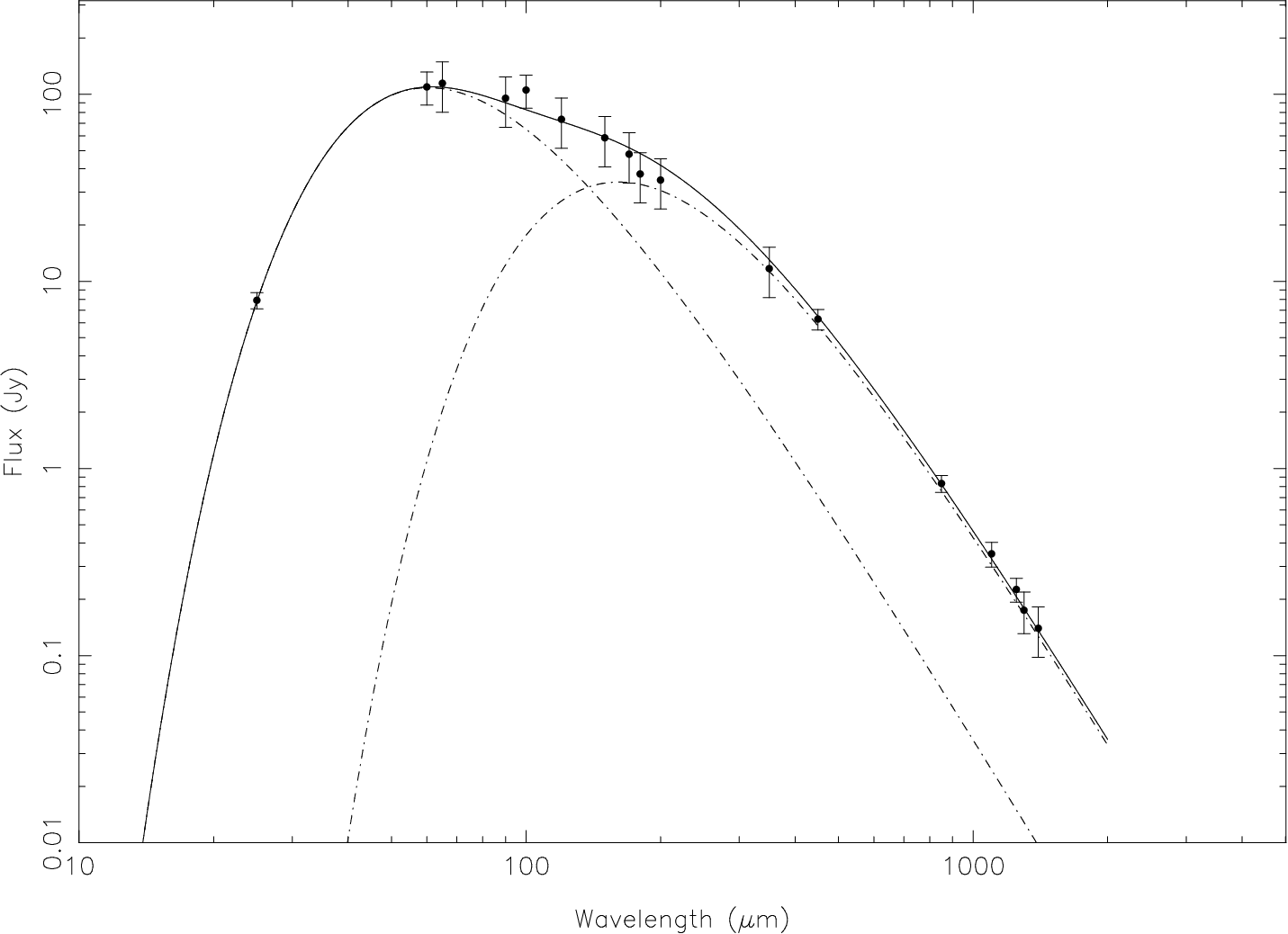,width=4.5cm,height=4cm,angle=-90}}\goodgap
 \subfigure[ngc7592: $T$(39,19), n=20
]{\psfig{file=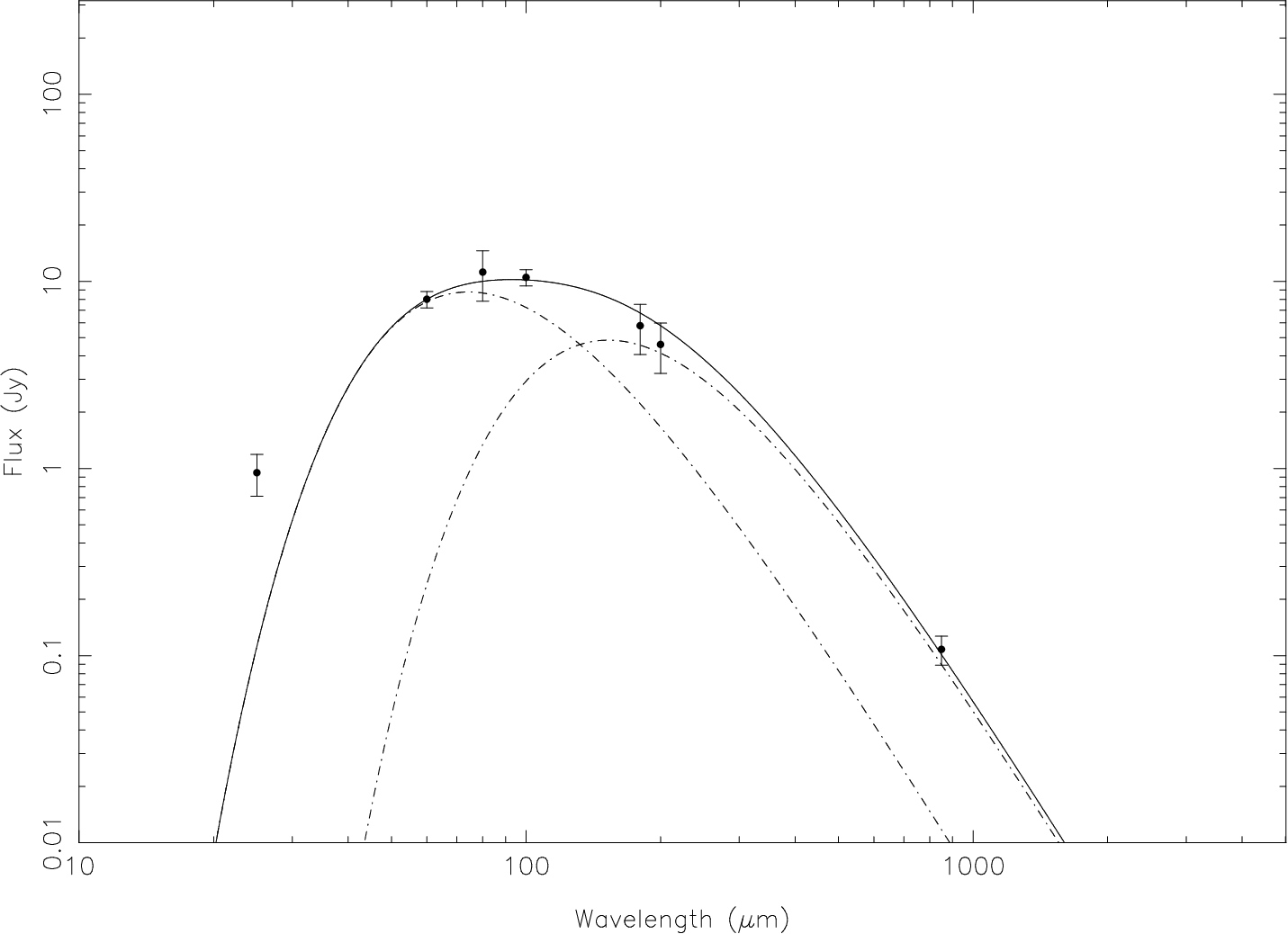,width=4.5cm,height=4cm,angle=-90}}\\
\caption{\label{2sedsF} Two-component SEDs assuming $\beta=2$ for IRAS
sample. The solid lines represent the composite 2-component SED while
the dot-dash lines show the warm and cold components. The caption to each figure gives the two temperatures plus the ratio of
cold:warm dust (n).}
\end{figure*}

The galaxies were re-fitted using Eqn.~\ref{2compSEDE} but now
assuming a fixed $\beta=2$. The results are given in
Table~\ref{beta2SEDT} and some examples of the fits are shown in
Fig.~\ref{2sedsF}.

\renewcommand{\baselinestretch}{1.0}

\renewcommand{\baselinestretch}{1.0}
\begin{table*}
\centering
\small
\caption{\label{beta2SEDT} Temperatures and dust masses for
two-component SEDs assuming $\beta=2$.}
\begin{tabular}{lcccccccccc}
\\[-2ex] 
\hline
\\[-2.5ex]
\multicolumn{1}{c}{(1)}&\multicolumn{1}{c}{(2)}&\multicolumn{1}{c}{(3)}&\multicolumn{1}{c}{(4)}&\multicolumn{1}{c}{(5)}&\multicolumn{1}{c}{(6)}&\multicolumn{1}{c}{(7)}&\multicolumn{1}{c}{(8)}&\multicolumn{1}{c}{(9)}&\multicolumn{1}{c}{(10)}&\multicolumn{1}{c}{(11)}\\
\multicolumn{1}{c}{Name}&\multicolumn{1}{c}{$T_w$}&\multicolumn{1}{c}{$T_c$}&\multicolumn{1}{c}{$\frac{N_c}{N_w}$}&\multicolumn{1}{c}{$M_{\rm{d2}}$}&\multicolumn{1}{c}{$R_{M}$}&\multicolumn{1}{c}{$L_{\rm{fir}}$}&\multicolumn{1}{c}{$R_L$}&\multicolumn{1}{c}{$G_{\rm{d(HI)}}$}&\multicolumn{1}{c}{$G_{\rm{d(H2)}}$}&\multicolumn{1}{c}{$G_{\rm{d}}$}\\
\multicolumn{1}{c}{}&\multicolumn{1}{c}{(K)}&\multicolumn{1}{c}{(K)}&\multicolumn{1}{c}{}&\multicolumn{1}{c}{(log $\rm{M_{\odot}}$)}&\multicolumn{1}{c}{}&\multicolumn{1}{c}{(log $\rm{L_{\odot}}$)}&\multicolumn{1}{c}{}&\multicolumn{1}{c}{}&\multicolumn{1}{c}{}&\multicolumn{1}{c}{}\\
\\[-2.5ex]
\hline
\\[-2.5ex]
UGC 903 & 45 & 21 & 91 & 7.37 & 1.95 & 10.34 & 1.17 & 141 & ... & ...\\
NGC 520 & 50 & 24 & 61 & 7.46 & 1.74 & 10.78 & 1.12 & 204 & 234 & 438\\
NGC 958 & 44 & 20 & 186 & 8.28 & 1.86 & 11.04 & 1.17 & 204 & 85 & 289\\
UGC 2369 & 36 & 21 & 6 & 8.08 & 2.04 & 11.34 & 1.02 & ... & 269 & ...\\
UGC 2403 & 50 & 22 & 100 & 7.57 & 2.04 & 10.71 & 1.17 & ... & ... & ...\\
UGC 2982 & 31 & 17 & 6 & 8.11 & 2.34 & 10.99 & 1.00 & 91 & 132 & 223\\
NGC 1614 & 38 & 20 & 3 & 7.78 & 2.14 & 11.32 & 1.00 & 58 & 174 & 232\\
NGC 1667 & 28 & 17 & 3 & 7.91 & 1.82 & 10.81 & 1.00 & 69 & 117 & 186\\
NGC 2623 & 50 & 27 & 18 & 7.59 & 1.58 & 11.36 & 1.07 & 46 & 240 & 286\\
NGC 2856 & 45 & 22 & 58 & 7.08 & 1.91 & 10.23 & 1.12 & 117 & ... & ...\\
NGC 2990 & 42 & 21 & 58 & 7.33 & 1.91 & 10.35 & 1.12 & 229 & ... & ...\\
UGC 5376 & 44 & 21 & 87 & 7.11 & 1.91 & 10.05 & 1.12 & 200 & ... & ...\\
NGC 3110 & 42 & 23 & 44 & 7.93 & 1.58 & 11.10 & 1.10 & ... & 257 &...\\
IR 1017+08 & 40 & 19 & 8 & 8.20 & 2.95 & 11.52 & 1.10 & ... & 151 & ...\\
IR 1056+24 & 40 & 26 & 6 & 8.14 & 1.51 & 11.79 & 1.02 & ... & 166 & ...\\
Arp 148 & 37 & 21 & 12 & 8.28 & 2.09 & 11.41 & 1.05 & ... & 85 & ...\\
MCG+00-29-023 & 36 & 20 & 13 & 7.99 & 2.14 & 11.04 & 1.05 & ... & ... & ...\\ 
IR 1211+03 & 49 & 24 &27 & 8.55 & 2.29 & 12.05 & 1.12 & ... & 123 & ...\\
NGC 4418 & 58 & 21 & 62 & 7.40 & 3.47 & 10.77 & 1.15 & 22 & 52 & 74 \\
UGC 8387 & 33 & 24 & 2 & 7.92 & 1.29 & 11.39 & 1.00 & ... & 148 & ...\\
Zw 247.020 & 39 & 25 & 6 & 7.50 & 1.51 & 11.04 & 1.02 & ... &... & ...\\
Zw 049.057 & 44 & 23 & 28 & 7.74 & 1.82 & 11.05 & 1.07 & ... & 65 & ...\\
1 Zw 107 & 39 & 20 & 7 & 8.23 & 2.63 & 11.56 & 1.02 & ... & 93 & ...\\
IR 1525+36 & 45 & 19 & 15 & 8.29 & 3.98 & 11.69 & 1.00 & ... & ... & ...\\
Arp 220 & 48 & 18 & 42 & 8.80 & 3.24 & 11.95 & 1.02 & 48 & 41 & 89\\
NGC 5962 & 31 & 18 & 11 & 7.49 & 2.19 & 10.26 & 1.17 & 98 & 120 & 218\\
NGC 6052 & 37 & 20 & 16 & 7.65 & 2.14 & 10.72 & 1.07 & 170 & 85 & 255\\
NGC 6181 & 31 & 19 & 7 & 7.47 & 1.86 & 10.39 & 1.05 & 191 & 195 & 386\\
NGC 7541 & 31 & 17 & 6 & 7.90 & 2.40 & 10.77 & 1.02 & 195 & 112 & 307\\
NGC 7592 & 39 & 19 & 20 & 8.13 & 2.75 & 11.14 & 1.05 & 87 & 129 & 216\\
NGC 7679 & 37 & 19 & 14 & 7.75 & 2.51 & 10.81 & 1.05 & 112 & ... & ...\\
NGC 7714 & 42 & 21 & 13 & 7.05 & 2.34 & 10.39 & 1.05 & 151 & 191 & 342\\
\\[-2.5ex]
\hline
\multicolumn{7}{c}{}&\multicolumn{1}{c}{mean}&\multicolumn{1}{c}{128}&\multicolumn{1}{c}{142}&\multicolumn{1}{c}{253}\\
\hline
\end{tabular}
\flushleft
\small (1) Name. (2) Warm temperature using
$\beta=2$. (3) Cold temperature using $\beta=2$. (4) Ratio of cold to
warm dust. (5) Dust mass calculated using the parameters in columns
(2--4). (6) Ratio of two-component mass to single temperature mass
from Dunne et al. (2000). (7)
FIR luminosity (40--1000$\mu$m) integrated under the 2 component
SED. (8) Ratio of this $L_{\rm{fir}}$ to that calculated assuming a
single temperature SED from Dunne et al. (2000). (9) H{\sc{i}} Gas-to-dust
ratio. (10) $\rm{H_2}$ gas-to-dust ratio. (11) Total gas-to-dust ratio
(sum of H{\sc{i}} and $\rm{H_2}$). Gas masses are those used in Dunne et
al. (2000).
\end{table*}

\subsubsection*{Gas and dust masses}
The dust mass has to be re-calculated if a colder component is
present, and will always be higher than if only one temperature is
assumed. The two-component dust mass is calculated as follows:

\begin{equation}
M_{d2} =
\frac{S_{850}\,D^{2}}{\kappa_{d(850)}}\times\left[\frac{N_c}{{\rm{B}}(850\mu
m,\,T_c)}+\frac{N_w}{{\rm{B}}(850\mu m,\,T_w)}\right] \label{coldmassE} 
\end{equation}

where $\kappa_{d(850)}$ is assumed to have the same value as in Paper
I (0.077 m$^2$ kg$^{-1}$). It is the temperature of the colder
component which is critical for an accurate determination of the dust
mass as it is this dust component which produces most of the emission
at 850$\mu$m. If $T_c$ has been overestimated then the dust mass will
be underestimated, in much the same way as if only a single
temperature had been assumed. The FIR luminosity can easily be
integrated under the new two-component SED and this and the dust
masses are given in Table~\ref{beta2SEDT}, along with the revised
gas-to-dust ratios. On average, the dust masses increase by a factor
$\sim 2$ compared to the single temperature estimates from Paper I. In
contrast, the values of $L_{\rm{fir}}$ do not increase by much at all
and this is simply because the long wavelength region, where the cold
component has modified the SED, does not contribute much to the
integrated energy. The gas-to-dust ratios are also lower than in our
earlier paper by a factor $\sim 2$, bringing them more into line with
the Galactic value of 226\footnote{The value quoted in Sodroski et
al. (1997) is 167 but we have had to scale their
$\kappa_d(240)=0.72$ m$^2$ kg$^{-1}$ to ours at 850$\mu$m. Using
$\beta=2$ the value of $\kappa_d(850)$ appropriate for Sodroski et
al. is 0.057m$^2$ kg$^{-1}$. Therefore their $G_d$ must be multiplied
by 0.077/0.057 to be consistent with ours, giving the value 226.}
(Sodroski et al. 1997). There is a slight tendency for the galaxies
with the largest angular sizes to have higher gas-to-dust ratios (NGC
520, NGC 958, NGC 6181, NGC 7541) which could be due to us missing
some extended dust emission outside the region mapped by SCUBA.

\renewcommand{\baselinestretch}{1.0}
\begin{figure}
\psfig{file=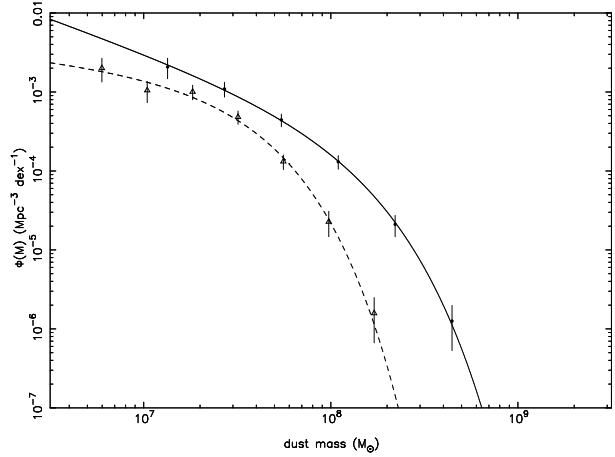,height=6cm,width=8.0cm,angle=-90}
\caption{\label{colddmfF} Dust mass functions. Solid line:-- `cold
dust mass' function, dust masses are from Table~\ref{beta2SEDT} and
from Paper I using a second cold component with
$T_{\rm{c}}=20\,\rm{K}$ and $\beta=2$. Dashed line -- dust masses
calculated from the 850$\mu$m fluxes using a single temperature as in
Paper I. Schechter parameters shown are $\alpha=-1.85$, 
$M_{\rm{d2}}^{\ast}=9.3\times10^{7}$ M$_{\odot}$ and $\phi^{\ast}=
4.9\times10^{-4}$ Mpc$^{-3}$ dex${-1}$ (solid); $\alpha=-1.23$,
$M_{\rm{d}}^{\ast}=2.5\times10^7$ M$_{\odot}$ and
$\phi^{\ast}=1.6\times10^{-3}$ Mpc$^{-3}$ dex$^{-1}$ (dashed).}
\end{figure}
For the galaxies without useful 450$\mu$m data, one can obtain
reasonable estimates of dust mass by fitting a two-component model
with $T_c$ and $\beta$ fixed at 20 K and 2. This is actually the
procedure used in Paper I to investigate the possibility of cold
dust. The dust mass function for the whole sample using the dust
masses derived in this paper and those `cold dust masses'
($M_{\rm{d2}}$) given in Table 4 of Paper I is
shown in Fig.~\ref{colddmfF}. This function could still
be an underestimate if there are large numbers of cold, dusty galaxies
which are not present in the {\em IRAS\/} sample.

\subsection{Dust heating}
\renewcommand{\baselinestretch}{1.0}
\begin{figure}
\psfig{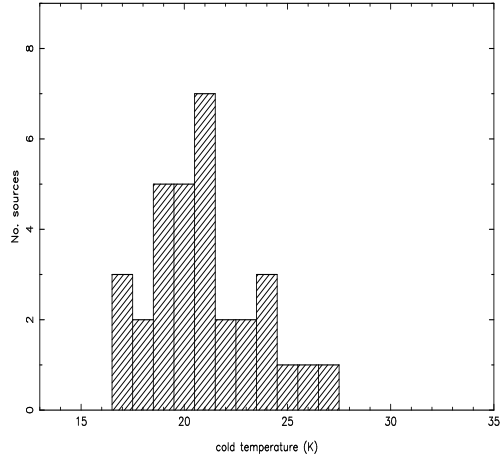}
\caption[Distribution of cold and warm temperatures using $\beta=2$.]{\label{temphist2F} Distribution of cold temperatures using $\beta=2$.}
\end{figure}

The new distribution of $T_c$ is shown in Fig.~\ref{temphist2F} and is
now much tighter than when $\beta$ was not fixed; it is also more
Gaussian with a single peak at 21 K. The mean is 20.9 K with
$\sigma=2.5$, and both the mean value and the small spread agrees with
the range of values for $T_c$ found in the literature (15 -- 25 K:-
Haas et al. 1998,2000; Frayer et al. 1999; Braine et al. 1997; Dumke et
al. 1997; Calzetti et al. 2000; Gu\'{e}lin et al. 1995; Sievers et
al. 1994; Neininger et al. 1996; Alton et al. 1998). It is also
interesting to note that alternative approaches to the SED inversion
problem presented by P\'{e}rez Garcia et al. (1998) and Hobson et
al. (1993) also require two or more temperature components and produce
similar temperatures. When observations are made with enough spatial
resolution, the SEDs of the central and outer regions of a galaxy can
be fitted separately. This often shows that the central regions
contain most of the warm dust at $T>30$ K while the outer disk regions
are dominated by the colder ($15-20$ K) dust, the total SED being the
sum of the inner and outer (Papadopoulos
\& Seaquist 1999; Haas et al. 1998; Dumke et al 1997; Braine et
al. 1997; Trewhella et al. 2000). This suggests that $N_c/N_w$ (or the
prominence of the cold component) is really telling us about the
relative importance of the disk/bulge dust components in these
galaxies (similar to the cirrus/starburst components in the models of
Rowan-Robinson \& Crawford 1989). The more active starburst galaxies
in the sample display conditions like the centres of normal galaxies
over a much larger region, so that the central warmer component
dominates the SED. The fact that the warm dust always out-shines the
cooler dust (per unit mass) is what makes it so difficult to
disentangle the cold component from the SED in cases where there is a
relatively large amount of warm dust. 

The equilibrium temperature of a dust grain immersed in a radiation
field can be expressed as a function of the interstellar radiation
field (ISRF) as (van der Hulst 1946; Disney et al. 1989)
\[
T_{\rm{eq}} \propto \rm{ISRF}^{1/5}
\]
This would lead one to expect variations in $T_c$ if some galaxies
(e.g. the more active starbursting ones) have higher ISRFs -- leading to
slightly warmer grain temperatures in their diffuse, cold
component. The range of $T_c$ found in these galaxies (17--27 K)
corresponds to a factor of $\sim 10$ in ISRF intensity, although it
may have been expected that the more active star forming galaxies in
the sample would have ISRFs many orders of magnitude greater than that
in the Milky Way. However, these active environments are also
localised and dusty, therefore a lot of the extra short wavelength
stellar radiation (UV) may be absorbed near to the source, producing
enhanced FIR radiation which is in fact the warm component. It is
generally believed that older stars contribute significantly to the
heating of the diffuse `cirrus' dust component as well as some
radiation from OB stars which leaks from the star forming molecular
clouds, although the relative importance of OB stars as a source of
heating is still not well determined (Bothun, Lonsdale \& Rice 1989;
Cox, Kr\"{u}gel \& Mezger 1986; Walterbos \& Greenawalt 1996;
Boulanger \& Perault 1988). If OB radiation leaked from sites of star
formation is not an important heating source for the colder component
then the distribution of older stars and diffuse dust must be
responsible for any changes in $T_c$. We will leave an investigation of the
details of the dust heating to a future paper. 

\renewcommand{\baselinestretch}{1.0}
\begin{figure}
\centerline{\psfig{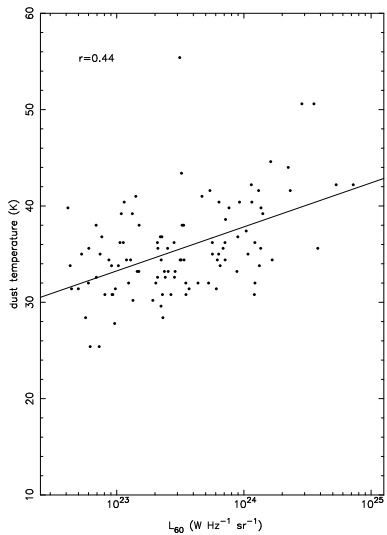}}
\caption{\label{L60TdF} 60$\mu$m luminosity correlated with dust
temperature (from isothermal models) in Paper I.}
\end{figure}

\renewcommand{\baselinestretch}{1.0}
\begin{figure*} 
\centerline{
 \subfigure[Warm temperature vs. $L_{60}$.]{\psfig{file=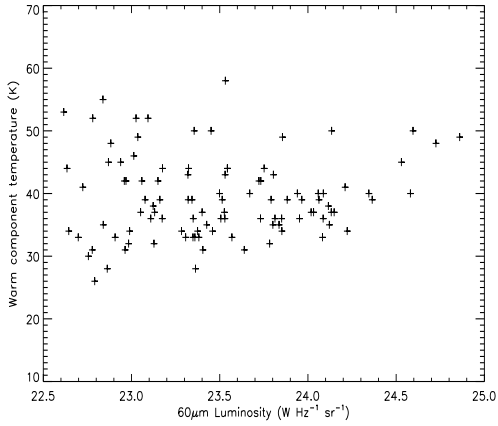,width=7cm,height=6cm}}
 \subfigure[Warm `energy' vs. $L_{60}$.]{\psfig{file=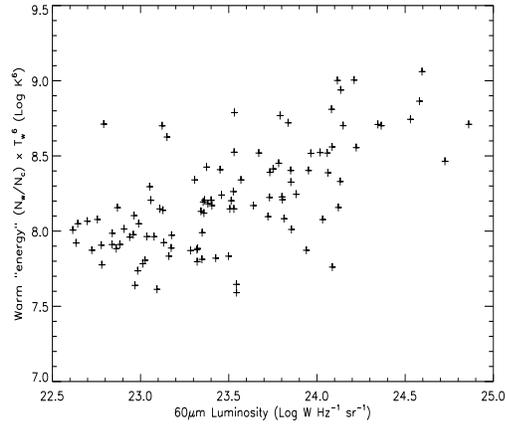,width=7cm,height=6cm}}}
\caption{\label{L60EwF} a) Using the warm
component temperature $T_w$, there is no longer a relationship with
$L_{60}$. b) Creating a warm `energy' by combining $T_w$ with the
ratio of cold/warm grains, there is now a correlation with $L_{60}$. }
\end{figure*}

\renewcommand{\baselinestretch}{1.0}
\begin{figure*}
\centerline{
 \subfigure[Dust mass vs.
$L_{60}$.]{\psfig{file=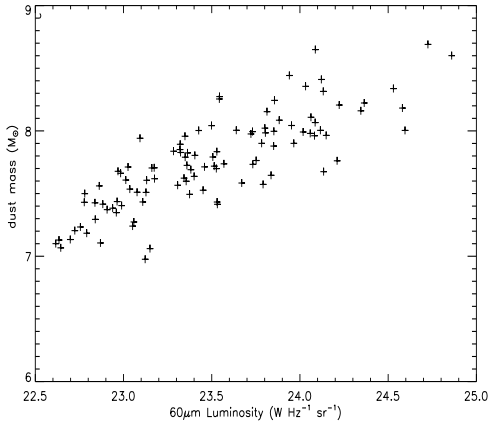,width=7cm,height=6cm}}
 \subfigure[Warm `luminosity' vs.
$L_{60}$.]{\psfig{file=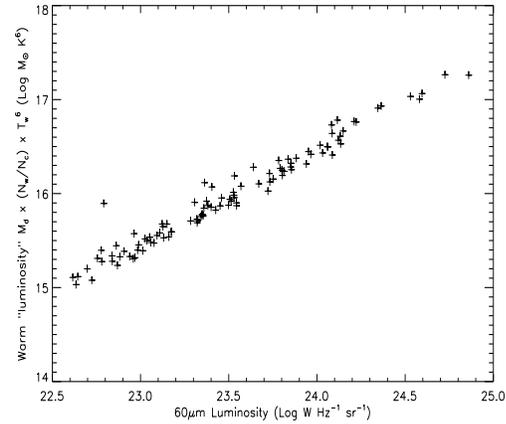,width=7cm,height=6cm}}}
\caption{\label{L60LwF} a) Dust mass (using the two-component
parameters) does correlate with $L_{60}$, but
the slope is less than unity meaning there must be some other
contribution to $L_{60}$ than mere size. b) Combining dust mass with
the warm `energy' to produce a distant dependent quantity (termed warm
`luminosity') gives a remarkably tight relationship with $L_{60}$
which does have a slope of unity.}
\end{figure*}

\renewcommand{\baselinestretch}{1.0}
\begin{figure*}
\centerline{
 \subfigure[Cold `energy'
vs. $L_{60}$]{\psfig{file=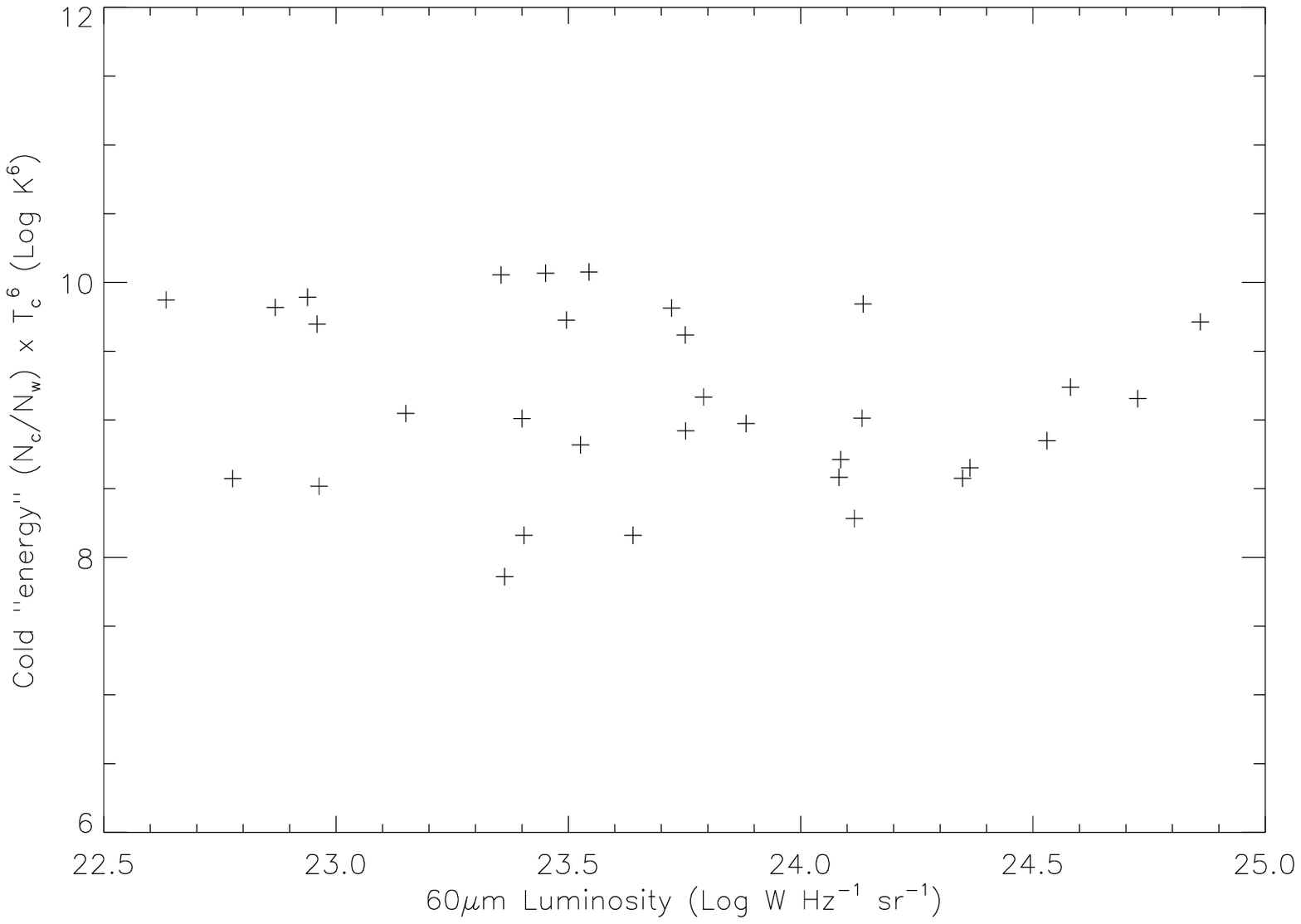,height=6cm,width=7cm}}
 \subfigure[Cold `luminosity'
vs. $L_{60}$.]{\psfig{file=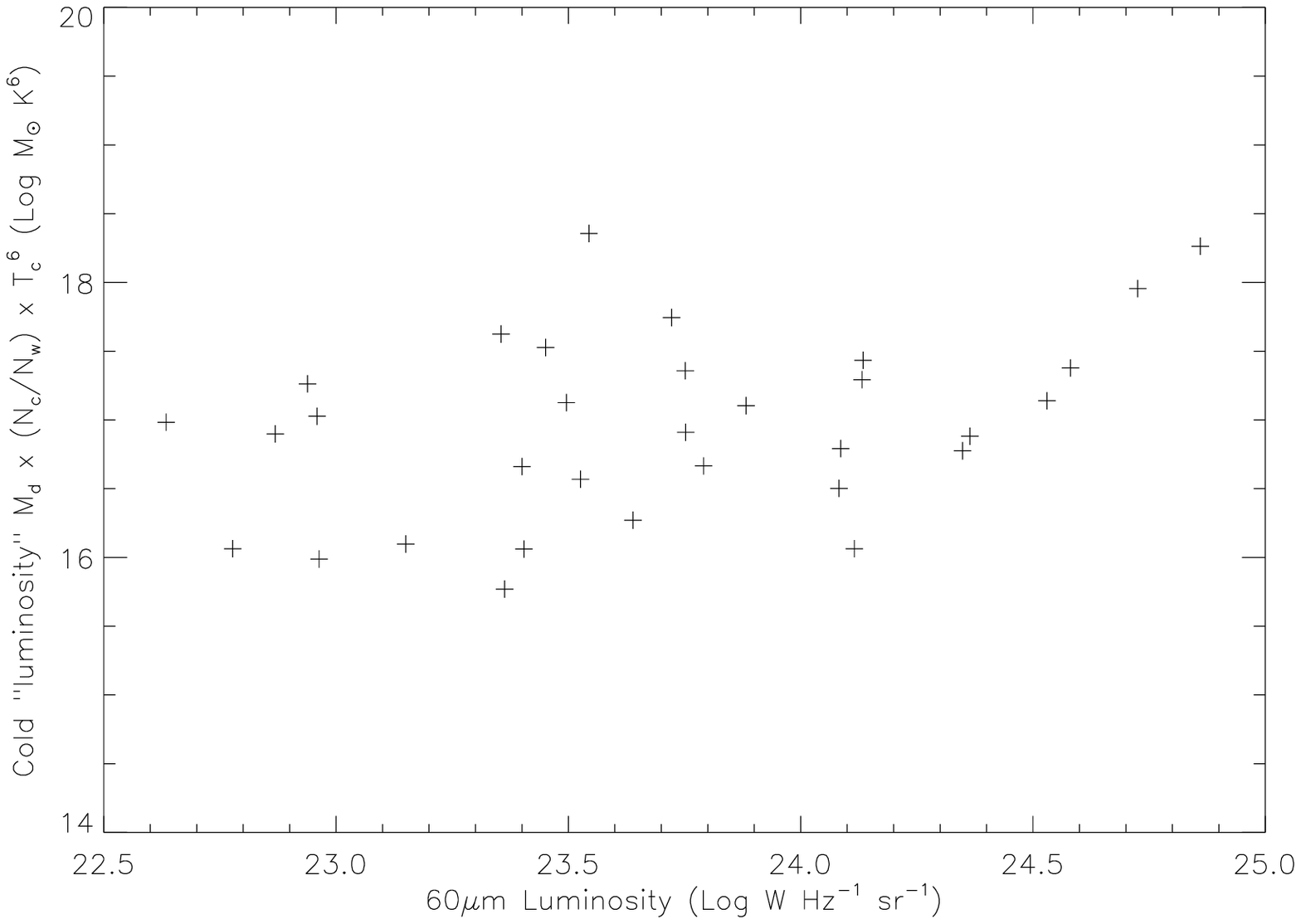,height=6cm,width=7cm}}}
\caption{\label{L60LcF} a) Creating an analogous cold `energy' using
$T_c$ and $N_c/N_w$ does not correlate at all with $L_{60}$ in
contrast to Fig.~\ref{L60EwF}b. b) Multiplying the cold `energy' by
dust mass to create the cold luminosity is only barely dependent on
$L_{60}$ via the relationship between dust mass and luminosity. This
highlights the significance of the strong relationship in
Fig.~\ref{L60LwF}b.}
\end{figure*}
 
\begin{table*}
\caption{\label{fitsT} Parameters for fits and correlations}
\begin{tabular}{ccccccc} 
\\[-2ex]
\hline
\\[-2.5ex]
\multicolumn{1}{c}{$y$}&\multicolumn{1}{c}{$x$}&\multicolumn{1}{c}{$r_{s}$}&\multicolumn{1}{c}{Prob}&\multicolumn{1}{c}{$\sigma$}&\multicolumn{2}{c}{linear
fit: $y=mx+c$}\\
\multicolumn{5}{l}{}&\multicolumn{1}{c}{$m$}&\multicolumn{1}{c}{$c$}\\
\\[-2.5ex]
\hline
\\[-2.5ex]
$T_{\rm{d}}$ & log $L_{60}$ & 0.44 & $2.5e-6$ & 4.7 & $4.57\pm 0.82$ &
$-71.9\pm 19.4$\\
$T_{\rm{w}}$ & log $L_{60}$ & $0.08$ & 0.42 & 0.8 & ..& ..\\
log [$N_w/N_c \times T_{\rm{w}}^6$] & log $L_{60}$ & $0.62$ & 2.1e-12 &
6.3 & $0.735\pm 0.068  $ & $-9.04\pm 1.25 $ \\
log $M_{\rm{d2}}$ & log $L_{60}$ & 0.83 & $2.7e-27$ & 8.4 & $0.727\pm 0.036$ &
$-9.31\pm 0.70 $\\
log [$M_{\rm{d2}}\times (N_w/N_c)T_w^6$] & log $L_{60}$ & 0.98 & 0.0 &
9.9 & $1.018\pm 0.014$ & $-7.95\pm 0.29$\\
log [$N_c/N_w \times T_{\rm{c}}^6$] & log $L_{60}$ & 0.16 & 0.40 & 0.8
& .. & ..\\
log [$M_{\rm{d2}}\times (N_c/N_w)T_c^6$] & log $L_{60}$ & 0.35 & 0.053 &
1.9 & .. & ..\\
\\[-2.5ex]
\hline
\\[-2.5ex]
\end{tabular}
\flushleft
\small (1) \& (2) Variables (3) Spearman rank correlation coefficient.
(4) Probability that $x$ and $y$ are unrelated. (5) Significance of
correlation in S.D. (6) Least-squares linear fit parameters.
\end{table*}

In Paper I we presented a relationship between dust temperature (as
fitted by an isothermal model) and 60$\mu$m luminosity
(Fig.~\ref{L60TdF}). Our explanation for this was that the position of
the 60$\mu$m flux on the Wien side of the Planck function makes it
very sensitive to dust temperature.  For a given mass of dust, hotter
galaxies would be easier to detect in a 60$\mu$m flux limited sample
and therefore a dependence of 60$\mu$m luminosity on dust temperature
is observed. We can now investigate this further using the extra
information we have about how the dust is split into warm and cool
components. The correlation coefficients and fits to the following
relationships are detailed in Table~\ref{fitsT}. We will now use the
whole data set of 104 galaxies. Those galaxies without useful
450$\mu$m data were fitted with a two-component model using fixed
$\beta=2$ and $T_c =20$ K (as described in Paper
I). Fig.~\ref{L60EwF}a shows that there is no correlation between the
fitted warm temperature and 60$\mu$m luminosity. The quantity which is
important now is the proportion of energy in the warm component, which
can be described roughly as $(N_c/N_w)^{-1}\times T_{w}^6$. This has a
quite a strong relationship with $L_{60}$ as displayed in
Fig.~\ref{L60EwF}b, although there is a lot of scatter. This
relationship is not linear, having a slope less than unity. It must be
remembered here that luminosity is a distant dependent quantity but
the relative amount of `warm energy' is not as it has not been
normalised in any way to the galaxy size. Is there a strong connection
between galaxy size (or dust content) and $L_{60}$? In Paper I we
argued that the link between dust temperature and $L_{60}$ was not a
function of galaxy size (i.e. dust temperature and dust mass were
unrelated, which is still the case here). However, dust mass and
$L_{60}$ are clearly related as shown in Fig.~\ref{L60LwF}a although
the relationship is not linear either (slope is less than unity) and
therefore 60$\mu$m luminosity would seem to depend on both the
quantity of dust in the galaxy and {\em the relative amount of energy
in the warm component\/}. If the `warm
energy' quantity from Fig.~\ref{L60EwF}b is multiplied by
$M_{\rm{d2}}$ to produce a distant dependent variable similar to
$L_{60}$ we find a remarkably tight correlation between this product,
the `warm luminosity', and the 60$\mu$m luminosity
(Fig.~\ref{L60LwF}b) and the slope of this correlation is unity. Thus
the typical FIR luminous objects detected by IRAS do not necessarily
have to have {\em vastly more dust\/} than lower luminosity ones, but
they do have to have a greater fraction of it heated to warmer
temperatures. We can repeat this comparison using the `cold energy'
$(N_c/N_w)\times T_c^6$ and `cold luminosity' $(N_c/N_w)\times T_c^6
\times M_{\rm{d2}}$. Neither of these quantities correlate with
$L_{60}$ as shown in Fig.~\ref{L60LcF}. This is an interesting point
because if there are galaxies which possess very dominant cold
components, even though they may have large quantities of dust, they
will not necessarily have large 60$\mu$m luminosities. This means they
may not be part of 60$\mu$m flux limited samples and hence represent a
population so far absent from our luminosity and dust mass
functions. This issue must await the completion of data analysis from
our optically selected sample, which should not be biased against
galaxies with large fractions of colder rather than warmer dust. 

\section{Conclusions}
We have used our own SCUBA 450$\mu$m data and fluxes from the
literature to produce better constraints on the FIR/submm SEDs of the
SLUGS {\em IRAS\/} sample. It was often not possible to find unique
decompositions of the SEDs using a series of modified Planck
functions, either cold temperatures and steep $\beta$ or higher
temperatures and lower $\beta$ being equally acceptable. However, by
the fact that the 450/850$\mu$m flux ratio is remarkably constant from
extreme objects like Arp220 to lower luminosity galaxies we concluded
the dust emissivity index $\beta$ could be constrained to be $\sim
2$. Using this value for $\beta$, a colder component is required in most
of the sample, but with a very large variation in its
contribution. The cold component has a similar temperature in all of
the galaxies ($T_{\rm{av}}=21$ K), in agreement with those determined
by many other authors in the literature. The dust masses derived using
the new temperatures are higher the previous estimates (which used a
single-temperature SED) by a factor $\sim 2$, and this brings the
gas-to-dust ratios into agreement with that of the Milky Way, and
other evolved spiral galaxies. The 60$\mu$m luminosity was found to depend on
{\em both} dust mass and the relative amount of energy in the warm
component, but it has no relationship with the cold component. 
Galaxies where the cold component is very dominant may not be well
represented in 60$\mu$m flux limited samples.

The support of PPARC is gratefully aknowledged by
L. Dunne, S. Eales. We also thank the JCMT staff especially Wayne Holland and
Iain Coulson for advice on 450$\mu$m calibration. We appreciate useful
discussions with Paul Alton.

\end{document}